\documentclass[a4, 12pt]{article}
\usepackage[a4paper,top=2cm,bottom=2cm,left=2cm,right=2cm]{geometry}
\usepackage{algorithm2e,amsmath,mathtools,amsthm,amssymb,authblk}
\usepackage{booktabs}
\usepackage{cellspace}
\usepackage{cite} 
\usepackage{color}
\usepackage{caption}
\usepackage[english]{babel}
\usepackage{enumerate,enumitem} 
\usepackage{epigraph}
\usepackage{float}
\usepackage{graphics,graphicx,latexsym,amsfonts} 
\usepackage{lineno}
\usepackage{multirow}
\usepackage{multicol}
\usepackage{pst-plot,pstricks}
\usepackage{picture}
\usepackage{subfig}
\usepackage{indentfirst} 
\usepackage[T1]{fontenc}
\def\gt{\textsc{gt}}
\def\gtt{\textsc{gt }}
\def\kcx{\textsc{kcx}}
\def\kcxx{\textsc{kcx }}

\def\mek{\textsc{mek}}
\def\mekk{\textsc{mek }}
\begin{document}

\title{Mr.Keynes and the...Complexity! \\ A suggested model for the \textit{General Theory}}

\author{Alessio Emanuele Biondo}
\affil{\small{Department of Economics and Business - University of Catania \\ Department of Business and Economics - TU Dortmund}}
\date{} 
\maketitle

\abstract{This paper presents a mathematical model of J.M. Keynes' \textit{The General Theory of Employment, Interest, and Money}. The main message of Keynes remains intact: it is the preference for liquidity, the marginal efficiency of capital, and the marginal propensity to consume that determine the level of aggregate income and employment, given any possible interest rate. All parts of the model are strictly referred to the original text.
\\

} 

\section{Introduction}

The \textit{General Theory} (\textsc{gt}, henceforth) begins with a clear statement of intentions (p.$3$): 
\begin{quotation}
	I have called this book the General Theory of Employment, Interest and Money, placing the emphasis on the prefix general. The object of such a title is to contrast the character of my arguments and conclusions with those of the classical theory of the subject, upon which I was brought up and which dominates the economic thought 
\end{quotation}
Since the day after its publication, in $1936$, an extensive range of contributions have enriched the scholarly debate. Some have been entitled by using either the word ``Keynes'' or the wording ``Mr.~Keynes and\ldots''. Many advanced comparisons with other economists or schools of economic thought. A first set of contributions adopts a biographical or broad intellectual perspective. Schumpeter (1946) offers a memorial portrait of Keynes as an economist and public intellectual, reconstructing the overall arc of his thought and situating his contribution within the history of economic analysis. Samuelson (1946) celebrates Keynes's legacy by underscoring the rupture that the \textit{General Theory} represented with respect to neoclassical orthodoxy, documenting its immediate and lasting influence on macroeconomic thinking. Skidelsky (2003) provides the definitive intellectual and political biography, tracing the evolution of Keynes's ideas from their philosophical roots in the \textit{Treatise on Probability} to their policy implications in the postwar international order. Dostaler (2010) pursues a comparable endeavour in the French-language tradition, reconstructing Keynes's theoretical and political engagements across their full European context. Moggridge (1976) offers a concise but authoritative biographical synthesis centred on the internal development of Keynes's thought across his major works. Davidson (2007) revisits Keynes's figure and theoretical legacy from a post-Keynesian standpoint, emphasising the foundational role of fundamental uncertainty and non-ergodicity in the original message of the \textit{General Theory}. Wood (1994) assembles and critically evaluates a wide body of secondary literature on Keynes as part of a comprehensive \textit{Critical Assessment} series. Minsky (1975) reinterprets Keynes through the lens of his own financial instability hypothesis, arguing that a correct reading of the \textit{General Theory} places endogenous financial fragility and cyclical crisis at the centre of capitalist dynamics, a reading subsequently extended in Minsky (1986).

A second subset of contributions draws explicit comparisons between Keynes and specific economists or theoretical traditions. Alexander (1940) juxtaposes Keynes and Marx, identifying convergences and divergences in their respective critiques of capitalism and their treatments of effective demand and the monetary circuit. Dillard (1948) situates the Keynesian revolution within the history of economic development, assessing its transformative implications for thinking about growth, accumulation, and stagnation. Dillard (1980) connects Keynes to the institutionalist tradition through the concept of a ``monetary theory of production,'' recovering often-neglected points of theoretical contact between Keynes and Veblen's approach to the money economy. Lawson and Lawson (1990) draw a parallel between Keynes, Veblen, and Kalecki in the context of financial system restructuring, arguing that all three provide complementary insights into the dynamics of financial instability and the political economy of reform. Robinson (1978) contrasts Keynes and Ricardo on the theory of distribution and effective demand, arguing that a genuinely Keynesian approach to surplus and accumulation requires a fundamental departure from the Ricardian problematic. Cutler et al.\ (2013) broaden the comparison to include Beveridge, situating Keynes within the wider landscape of twentieth-century British social and welfare thought and examining the political tensions between Keynesian demand management and the architecture of the welfare state. Buchanan et al.\ (1978) engage with Keynes from the perspective of public choice theory, arguing that the discretionary fiscal logic implicit in the \textit{General Theory} carries damaging institutional consequences for democratic governance and fiscal constitutions, as Keynesian legitimation of deficit spending removes the discipline of balanced-budget rules while assuming an implausible degree of technocratic insulation from political pressures.

A third group concerns the critical reception of the \textit{General Theory} and the long-running debate over its correct interpretation. Pigou (1936) constitutes the first major response from the leading representative of neoclassical orthodoxy, contesting Keynes's assumptions regarding the labour market and arguing that the \textit{General Theory} fails to establish a logically coherent account of involuntary unemployment distinct from frictional phenomena. Harrod (1937) defends the compatibility of the \textit{General Theory} with received theory, anticipating the neoclassical synthesis by framing Keynes's contribution as a generalisation rather than a repudiation of classical economics and its treatment of the interest rate. Hicks (1937) proposes the celebrated IS--LM formalisation, which would become the canonical, although non unanimously accepted, vehicle for the textbook transmission of Keynesian ideas and the framework within which most of the subsequent policy debate was conducted. Leijonhufvud (1967) decisively overturns this reading, arguing that the economics of the Keynesians diverges profoundly from the economics of Keynes himself, and that the IS--LM framework systematically suppresses the original disequilibrium message of the \textit{General Theory} by substituting a comparative-statics apparatus for Keynes's sequential, monetary analysis. Chick (1983) provides a systematic and chapter-by-chapter reconsideration of the \textit{General Theory}, arguing that the standard IS--LM interpretation trivialises Keynes's analysis by separating the monetary from the real sector in a way that contradicts Keynes's own insistence on the monetary character of production; she develops an alternative reading grounded in the logic of a monetary production economy and the sequential nature of economic processes. Patinkin (1990) returns to and deepens the debate over competing interpretations of the \textit{General Theory}, examining the role of disequilibrium, the real-balance effect, and the conditions under which involuntary unemployment can be grounded in a theoretically coherent framework. Krugman (2011) intervenes in the post-crisis debate by reasserting the relevance of Keynes against the macroeconomic mainstream, arguing that the \textit{General Theory} anticipates core insights that modern DSGE models systematically suppress by construction. Blinder (1987) offers a comparative assessment of scientific progress in the Keynesian versus Lucasian research programmes, concluding on the basis of the empirical record that the new classical revolution represented a detour rather than an advance. Worswick et al.\ (1983) collect contributions that reassess the pertinence of Keynes for the contemporary world, covering monetary policy, fiscal multipliers, open-economy complications, and the enduring macroeconomic legacy of the \textit{General Theory}.

Robinson (1964a) poses the counterfactual question of whether a \textit{General Theory} could have emerged without Keynes as an individual, exploring the extent to which the intellectual conditions of the 1930s made such a contribution historically overdetermined or, conversely, contingent on Keynes's unique combination of theoretical creativity and institutional position. Robinson (1964b) offers a detailed intellectual profile of Keynes in the same volume, emphasising the historical specificity of the Keynesian revolution and the diversity of influences that shaped it. Leith and Patinkin (1977) examine the relationship between Keynes, the Cambridge intellectual environment, and the genesis of the \textit{General Theory}, shedding light on the debates and correspondences that shaped the final text and the role of Cambridge as an institutional crucible. Kahn et al.\ (1984) reconstructs the intellectual process and the discussions of the Cambridge ``Circus'' that accompanied and stimulated the writing of the \textit{General Theory}, documenting the collaborative dimension of its elaboration and the role of Richard Kahn in the development of the multiplier.

Others focused on methodology. Carabelli (1988) is the seminal monograph on Keynes's method, demonstrating that his epistemological approach, grounded in a conception of probability inherited from the \textit{Treatise on Probability} and in the notion of irreducible uncertainty, is inseparable from his substantive theoretical results and cannot be reduced to a merely technical apparatus superimposed on otherwise conventional economic reasoning. Lawson and Pesaran (1985) bring together contributions that scrutinise the methodological presuppositions of Keynesian economics, engaging with the debate between formalism, realism, and ontological commitment in economic theorising, and examining the extent to which Keynes's framework requires a distinctive philosophical foundation. Kregel (1976) analyses the modelling strategies of Keynes and the post-Keynesians when confronted with the pervasiveness of uncertainty, distinguishing between static, shifting-equilibrium, and stochastic approaches and arguing that the post-Keynesian tradition is methodologically more faithful to the spirit of the \textit{General Theory} than the neoclassical synthesis.

The influence of Keynesian economics on specific topics or policy issues is discussed across a further set of contributions. Viner (1936) constitutes one of the earliest critical responses to the \textit{General Theory}, contesting Keynes's account of the causes of unemployment and arguing that the theory overstates the role of aggregate demand while underemphasising supply-side mechanisms and the classical adjustment process. Leontief (1936) identifies the ``homogeneity postulate'' as the fundamental assumption distinguishing Keynes from orthodox theory: Keynes, he argues, treats as an independent behavioural datum what is, from the standpoint of general equilibrium theory, properly a derived result, thereby elevating a questionable assumption to the status of a foundational axiom. Robertson (1938) engages in a precise and intellectually consequential exchange with Keynes over the ``finance motive'' for holding liquidity and its implications for the theory of interest, the loanable funds debate, and the relationship between saving, investment, and the money supply. Crotty (1983) examines Keynes's views on capital flight and the regulation of international capital movements, arguing that the \textit{General Theory} contains an underappreciated political economy of financial openness with direct implications for contemporary debates on capital controls and international monetary reform. Moggridge and Howson (1974) reconstruct the evolution of Keynes's thinking on monetary policy over nearly four decades, demonstrating continuity between his early writings and the \textit{General Theory} while documenting the shifting emphasis across different phases of his intellectual development, from the \textit{Tract on Monetary Reform} through the \textit{Treatise on Money} to the \textit{General Theory} itself. Leijonhufvud (1968) examines the conditions of effectiveness of monetary policy within the Keynesian framework, distinguishing the liquidity trap as a structural feature of asset markets characterised by conventional expectations from its dismissal as a mere and empirically irrelevant limiting case. de Carvalho (2015) argues that all of the major theoretical innovations of the \textit{General Theory} are ultimately derivable from Keynes's analysis of defensive behaviour under fundamental uncertainty, offering a unified interpretive framework that counters the mechanistic and fiscalist readings of hydraulic Keynesianism and restores the centrality of agent behaviour under non-measurable uncertainty. Dow and Hillard (1995) assemble contributions exploring the relationship between knowledge, uncertainty, and Keynesian analysis, recovering the epistemological dimensions of Keynes's economics that standard macroeconomic readings systematically neglect. Tobin (1977) defends the intellectual vitality of Keynes in the midst of the monetarist offensive of the 1970s, identifying four core propositions of Keynesian macroeconomics, concerning aggregate demand, involuntary unemployment, wage flexibility, and the role of money in economic fluctuations, that he argues remain empirically compelling and theoretically defensible. Nell (1998) develops a Keynesian theory of transformational growth, extending the framework of the \textit{General Theory} in an evolutionary and Sraffian direction: he argues that the theory of transformational growth replaces equilibrium with history, that the role of markets is not to allocate given resources but to generate and select innovations, and that the resulting framework makes possible a revised Keynesian approach to monetary circulation grounded in classical foundations.

Also mathematical formalisations of Keynes's system have been previously proposed, as in Meade (1937), where an equilibrium-driven framework offers the first explicit mathematical formalisation of the Keynesian system, expressing the relationships among income, consumption, investment, the interest rate, and the money supply as a set of simultaneous equations. While remaining within a comparative-statics framework that presupposes market-clearing equilibrium, Meade's model anticipates the structure of the IS--LM apparatus subsequently developed by Hicks (1937) and constitutes a foundational reference point for all subsequent formal renditions of the \textit{General Theory}'s analytical core.

The intersection between Keynesian economics and the theory of complexity has itself attracted scholarly attention. Colander (2010) argues that Keynes's method was fundamentally oriented towards the analysis of complex systems, and that the progressive displacement of this method by the formalism of neoclassical economics, driven by academic incentive structures rather than scientific progress, has impoverished the training of economists and their capacity to address real-world macroeconomic problems. The relevance of complexity for understanding financial instability is further explored in Sornette (2009), who applies concepts and methods from statistical physics and complex systems theory to the analysis of stock market crashes, thus investigating from a perspective that resonates directly with Keynes's own emphasis on the destabilising role of financial markets and herd behaviour. Pasinetti (1997) contributes to the reassessment of the \textit{General Theory} by examining Keynes's concept of the marginal efficiency of capital in Book~IV, distinguishing between the autonomous component of investment, driven by long-term expectations and animal spirits, and its endogenous component, subject to formal economic calculus, and arguing that this distinction carries lasting implications for investment theory and for understanding the inducement to invest under uncertainty.

The present paper presents a mathematical model of the \gt, in which the sequentiality of interactions among all participants living in the economic system is shown to represent the distinctive feature of the Keynesian revolution. An operative model of the \gtt that aligns as closely as possible to the original text can provide meaningful opportunities in both theoretical and applied macroeconomic analysis.

In order to underline that the revolutionary perspective of Keynes lies in the conception of the economic system and not in the heterodoxy of methodologies, the present model describes agents as the orthodoxy of economics prescribes: all optimize individual target functions with respect to individual constraints. Keynes himself referred very often to utility and production functions in the \gt\footnote{This is not indispensable. An integrally behavioural approach is possible, yielding same conclusions.}: the main motivation here is to find theoretical underpinnings in his original text and build a consistent model\footnote{In principle, many alternative formulations are possible, of course.}.

The distinguishing feature of the Keynesian approach is the sequentiality of actions, which is the most fundamental ingredient of the complexity. The Keynes-CompleXity-model here proposed (\kcx, henceforth) is \textit{micro-designed}, thus letting the emergence of macro features be a result of agents' interactions (Epstein, 1999; Axtell, 2007). Complexity reveals to be crucial in understanding macroeconomics, as early understood by Keynes (1936) and Majorana (1942) and discussed in Kirman (2011) and Ladyman et al.\ (2013), among others. A complex environment is a configuration where the interaction among individual elements generates emergent aggregate outcomes that qualitatively differ from the features of its constituents, as spontaneous self-organized structures at different layers of a hierarchical configuration (Gallegati and Richiardi, 2009). It is, then, a matter of sequentiality, since individuals do not act simultaneously. Without coordination, consumers and firms maximize their own objective functions --- under related constraints --- on the basis of the perceived context. Thus, \textit{animal spirits} and subjective probabilities define a state of \textit{true} uncertainty, in the sense of Knight (1921). Given that the sequence of actions is not predetermined, precise numeric predictions cannot be meaningful, since \guillemotleft the evidence is insufficient to establish a probability\guillemotright and possible consequences of variations \guillemotleft are not even orderable\guillemotright (Hicks, 1979, pp.~113, 115). Macroeconomic systems are thus \textit{non-ergodic}, characterized by the plurality of choices belonging to different time steps, each unpredictably influenced by the previous and influencing the following ones.

If agents act simultaneously a permanent equilibrium can be reached. Then, only ``ad hoc'' shocks can explain fluctuations and the economy can be described by means of a representative agent (see Hartley, 1997, on this methodology, widely adopted in the mainstream macroeconomics approach). Contrariwise, the \gtt describes the economy as a dynamics of forces resulting by all individual decision rules and the \kcxx model follows this approach, along the lines traced in Kirman (1992), Akerlof (2002), Leijonhufvud (2006), Delli Gatti et al.\ (2008), Gaffeo et al.\ (2008), Delli Gatti et al.\ (2011), among others. From this perspective, agent-based models have a distinct advantage compared to \textsc{dsge} models, which try to account for heterogeneity while remaining substantially simultaneous-equilibrium devices, as described in Massaro (2013) and Christiano et al.\ (2018). Agent-based models have been widely adopted in macroeconomics, as in Delli Gatti et al.\ (2005, 2011), Dosi et al.\ (2006, 2008, 2010), Cincotti et al.\ (2010, 2012), Raberto et al.\ (2014), Salle et al.\ (2013) and Salle (2015), Wolf et al.\ (2013), Seppecher and Salle (2015), Ashraf et al.\ (2016, 2017), Dawid et al.\ (2018), among others. A comprehensive survey of macroeconomic agent-based models (\textsc{mabm}) is Dawid and Delli Gatti (2018).

With the purpose of illustrating the utmost importance of sequentiality, the first set of presented results reports the case of a population of identical individuals with perfectly homogeneous attributes, showing that they may --- however --- yield different results among themselves. The reason is simple: once a player moves, the second one's opportunities are unavoidably restricted upon the former's choice. The second series of results have been obtained by extensive Monte Carlo simulations with heterogeneous agents.

The paper is organized as follows: section two presents the model; section three shows results; section four presents some conclusive remarks and the forthcoming research lines.

\section{The \textsc{kcx} model}

The modelled society is populated by $n_p$ persons, characterized as consumers (workers) and entrepreneurs (each owing one of $n_f$ firms). The model is described by reporting original words written by Keynes in the \gt, since <<it may be useful to make clear which elements in the economic system we usually take as given, which are the independent variables of our system and which are the dependent variables.>> (\gt, p.$245$)

\subsection{Exogenous variables and the social context} \label{exogenous-variables}

<<We take as given the existing skill and quantity of available labour, the existing quality and quantity of available equipment, the existing technique, the degree of competition, the tastes and habits of the consumer, the disutility of different intensities of labour and of the activities of supervision and organisation, as well as the social structure including the forces, other than our variables set forth below, which determine the distribution of the national income. [...] This does not mean that we assume these factors to be constant; but merely that, in this place and context, we are not considering or taking into account the effects and consequences of changes in them.>> (\gt, p.$245$)

Such exogenous variables are then kept constant exclusively in order to allow the treatment of the model, which should be entirely re-computed again and again for all their possible values. Stated conditions have been here modelled as follows: 
\begin{itemize}
\item[-]  First, the population size ($n_p$) and skills of persons, either workers or entrepreneurs, are initially assigned (i.e., both quality and quantity of available labour are given).
\item[-] Secondly, the technology of production is given, although possibly heterogeneous among firms. Firms are distinguished between capital-goods producers (k-firms, henceforth) and consumption-goods producers (c-firms, henceforth). 
\item[-]  Third, the set of goods is defined as $G=G^k\cup G^c$, with $G^k=\{x^k_1,\ldots,x^k_{n_k}\}$ and $G^c=\{x^c_1,\ldots,x^c_{n_c}\}$. Hence, $|G^k|=n_k$, $|G^c|=n_c$, and $|G|=n_g=n_k+n_c$. Each good is produced by one firm only, so that the number of k-firms equals $n_k$ and the number of c-firms equals $n_c$. Each good has an intrinsic value, $v_{g_h}$, used in individual functions as detailed later. Consumption goods receive values from persons' preference vectors, whereas k-goods receive values from c-firms' capital-good preference vectors. In the implementation, the value of each good is computed as the corresponding cross-agent mean and is then used to determine the market power of its producer.

\item[-] Fourth, the number of firms and their market power do not vary in time. 
\item[-] Fifth, preferences of persons are individually assigned at the beginning, i.e., each person has an utility function..
\item[-] Sixth, the societal composition is fixed: workers remain workers and entrepreneurs (i.e., firms holders) remain entrepreneurs; further, the wage-unit is given.
\item[-] Seventh, all production functions have the same curvature for all the relevant employment levels (\gt, p.$246$). 
\item[-] Eighth, the labour supply function is supposed to be known (\textit{ibidem}). 
\end{itemize}

\subsection{Explanatory variables} \label{explanatory-variables}
<<Our independent variables are, in the first instance, the propensity to consume, the schedule of the marginal efficiency of capital and the rate of interest, though, as we have already seen, these are capable of further analysis>> (\gt, p.245). The three explanatory variables are, then, intended as an aggregation of individual behaviours, with the exception of the interest rate. 
<<Thus we can sometimes regard our ultimate independent variables as consisting of (1) the three fundamental psychological factors, namely, the psychological propensity to consume, the psychological attitude to liquidity and the psychological expectation of future yield from capital assets, (2) the wage-unit as determined by the bargains reached between employers and employed, and (3) the quantity of money as determined by the action of the central bank; so that, if we take as given the factors specified above, these variables determine the national income (or dividend) and the quantity of employment>> (\gt, pp.$246$-$7$). The \kcxx model moves from an initial level of the interest rate, $r_0$, in order to explain, on individual basis, the three aggregate \textit{fundamental psychological factors}, i.e., the liquidity preference, the decision to invest and the consumption choice.

\subsubsection{The Liquidity Preference}

\noindent The liquidity preference of a person <<[...] is given by a schedule of the amounts of his resources, valued in terms of money or of wage-units, which he will wish to retain in the form of money in different sets of circumstances (\gt, p.$166$). [...] Liquidity-preference is a potentiality or functional tendency, which fixes the quantity of money which the public will hold when the rate of interest is given; so that if $r$ is the rate of interest, $M$ the quantity of money and $L$ the function of liquidity-preference, we have $M = L(r)$. This is where, and how, the quantity of money enters into the economic scheme.>> (\gt, p.$168$). Keynes did not consider the interest rate as the remuneration of saving: it is, rather, one of the elements determining which form to give to the liquidity exceeding the consumption requirements: <<why should anyone prefer to hold his wealth in a form which yields little or no interest to holding it in a form which yields interest [...] ?>> (\textit{ibidem}). Therefore, the interest rate is the price which equates the desire to have liquid money in hands, given the available quantity of money. <<There is, however, a necessary condition failing which the existence of a liquidity-preference for money as a means of holding wealth could not exist. This necessary condition is the existence of uncertainty as to the future of the rate of interest, i.e. as to the complex of rates of interest for varying maturities which will rule at future dates>> (\textit{ibidem}). Then, in a world of certainty, where all returns opportunities at any time are known and, consequentially, the full employment had been reached, the Quantity Theory of Money would hold (\gt, p.$209$) whereas, in an uncertain world, different expectations for the future value of the interest rate induce different liquidity preferences.

In the \kcxx model, the individual function describing the liquidity preference of each agent $i$, namely $L_{2 \, i,t}(r_t)\equiv\,E[r_{i,t+1}]$, determines the expected interest rate as a function of the current interest rate and depicts the \textit{state of confidence}, i.e., <<a matter to which practical men always pay the closest and most anxious attention>> (\gt, p.$148$):
\begin{equation}
L_{2 \, i,t}(r_t)= (r_t \pm \varrho_{i,t})
	\label{liq-pref}
\end{equation}
In the homogeneous case, all persons are assigned the same value $L_{2\,i,t}=r_t+\bar{\varrho}$. In the heterogeneous case, individual liquidity preferences are drawn from a normal distribution centred on $r_t+\bar{\varrho}$, truncated at zero. This value also determines the monetary exponent of the consumption problem, with $z_{i,M}$ increasing in $L_{2\,i,t}$ and bounded in the implementation to avoid degenerate consumption shares. Then, <<who believes that future rates of interest will be above the rates assumed by the market, has a reason for keeping actual liquid cash, whilst the individual who differs from the market in the other direction will have a motive for borrowing money for short periods in order to purchase debts of longer term>> (\gt, p.$170$). Neither a formal treatment of the financial market nor a specification on the term-structure of interest rates have been included in the \gt. Thus, the portfolio analysis is maximally simplified also in the \kcxx model: saving (i.e., money not used for consumption) is allocated either in liquid money or bonds, giving a yield equal to the interest rate, if $L_{2 \, i,t}>r_t$ or $L_{2 \, i,t}<r_t$, respectively. Indeed,  <<[...] uncertainty as to the future course of the rate of interest is the sole intelligible explanation of the type of liquidity-preference $L_2$ which leads to the holding of cash $M_2$>> (\gt, p.$201$).

\subsubsection{The Marginal Propensity to Consume} 
<<Consumption --to repeat the obvious-- is the sole end and object of all economic activity. Opportunities for employment are necessarily limited by the extent of aggregate demand. Aggregate demand can be derived only from present consumption or from present provision for future consumption>> (\gt, p.$104$). Each agent chooses the part of money for present consumption and the part devoted to command over future one (p.$166$). The other part is allocated to more satisfactory rewards: either invested (in real or financial activities), or retained if the possession of liquidity is considered preferable. 

All possible assets, ranging from money to goods, have a potential return based on three ``attributes'': the yield, i.e., the positive effect of having that good; the carrying/maintenance cost, i.e., the set of costs associated to it; the liquidity premium, i.e., the potential convenience given by the power of disposal, net of costs (\gt, pp.$225$-$226$):
\begin{quotation}
Let us consider what the various commodity-rates of interest over a period of (say) a year are likely to be for different types of assets. Since we are taking each commodity in turn as the standard, the returns on each commodity must be reckoned in this context as being measured in terms of itself.
There are three attributes which different types of assets possess in different degrees; namely, as follows: (i) Some assets produce a yield or output $q$, measured in terms of themselves, by assisting some process of production or supplying services to a consumer.
(ii) Most assets, except money, suffer some wastage or involve some cost through the mere passage of time (apart from any change in their relative value), irrespective of their being used to produce a yield; i.e. they involve a carrying cost $c$ measured in terms of themselves. It does not matter for our present purpose exactly where we draw the line between the costs which we deduct before calculating $q$ and those which we include in $c$, since in what follows we shall be exclusively concerned with $q-c$.
(iii) Finally, the power of disposal over an asset during a period may offer a potential convenience or security, which is not equal for assets of different kinds, though the assets themselves are of equal initial value. [...] The amount (measured in terms of itself) which they are willing to pay for the potential convenience or security given by this power of disposal (exclusive of yield or carrying cost attaching to the asset), we shall call its liquidity-premium $l$. 
It follows that the total return expected from the ownership of an asset over a period is equal to its yield minus its carrying cost plus its liquidity-premium, i.e. to $q-c+l$. That is to say, $q-c+l$ is the own-rate of interest of any commodity, where $q$, $c$ and $l$ are measured in terms of itself as the standard.
\end{quotation}

The \kcxx model thoroughly implements the cited rationale. At each at time step $t$, each person $i$ maximizes a utility function, which positively depends on individual arguments: money, $m_{i,t}$, quantities of goods, $x_{i,j,t}$, and leisure time, $\mathcal{L}_{i,t}$, which are defined as having heterogeneous attributes, such that chosen values can be considered as the result of personal computations, i.e., $z_{i,M}=z(q_{i,M}-c_{i,M}+l_{i,M})$ for money, $z_{i,C}=z(q_{i,C}-c_{i,C}+l_{i,C})$ for consumption as a whole, and $z_{i,\mathcal{L}}=z(q_{i,\mathcal{L}}-c_{i,\mathcal{L}}+l_{i,\mathcal{L}})$ for labour. Since the disutility of labour is considered as given (see above), $z_{i,\mathcal{L}}=\bar{z}_\mathcal{L}$, $\forall i$. The coefficient for money has been defined as a direct function of the individual liquidity preference, i.e., $z_{i,M}=f(L_{2 \, i,t})$. Since it is assumed that $\sum z_{i,\bullet}=1$, then $z_{i,C}=1-z_{i,M}-z_{i,\mathcal{L}}$. 

Single c-goods $j$ have their heterogeneous attributes for each person $i$ as well, opportunely intended as $v_{i,j}=v(q_{i,j}-c_{i,j}+l_{i,j})$, all measured in psychological units by each person. Values are actually assigned as random variables drawn from the interval $(0,1)$ and, for each person $i$, it is $\sum_{j} v_{i,j}=1$. The value of each good is obtained by summing all individual values referred to it, i.e., $v_g=\sum_{i=1}^{n_p}v_{i,g}$, $\forall g \in [n_{k}+1, ... , n_{k}+n_{c}]$.

Optimal quantities of money and, consumption, and working time are obtained by each agent $i$ by maximizing the individual utility subject to the individual budget constraint:
\begin{equation}
\text{max} \,\,\,\, U_{i,t}=m_{i,t}^{\;\; z_{i,M}}\left[ \prod_{x_j \in G^c}x^{v_{i,j}}_{i,j,t}\right]^{z_{i,C}} \negthickspace \negthickspace \negthickspace \mathcal{L}_{i,t}^{\;\; z_{i,\mathcal{L}}} \quad \quad \text{s.t.} \quad \quad w(\bar{N} - \mathcal{L}_{i,t}) = \sum_{x \in G^c} p_j x_{i,j,t}+m_{i,t}
	\label{max-utility}
\end{equation}
where earned income, obtained by multiplying the wage unit $w$ by the individual labour supply, $N_{i,t}=(\bar{N} - \mathcal{L}_{i,t})$, sustains the monetary allocation (in liquid form, $m_{i,t}$, or in purchases of chosen consumption goods, $\sum_{x \in G^c_{i,t}} p_jx_{i,j,t}$). The Cobb-Douglas functional form, with $z_{i,M}+z_{i,C}+z_{i,\mathcal{L}}=1$, has been chosen because of its tractability in calculations. Many alternative approaches are, indeed, possible. The vast majority of models in agent-based literature, briefly recalled in the introduction, adopt random variables or stochastic functions, in consumption and/or in production. In the implementation used for the simulations, the wage is treated as an exogenous contextual variable, consistently with the approach of the \gt. Individual optimization is therefore performed conditional on a given wage level, so that no labour supply schedule is derived: a proper labour supply curve would require solving the same optimization problem across all admissible wage levels, which is beyond the scope of the present model. Employment is thus determined entirely on the demand side: firms demand labour according to their expected production and available factors, and workers are hired sequentially. Once income has been distributed, each person allocates positive net income between consumption and retained money according to the Cobb-Douglas shares implied by $z_{i,C}$ and $z_{i,\mathcal{L}}$.

Conditional on received net income, the implemented allocation rules for agent $i$ are:
\begin{equation}
	N_{i,t}^{*}=(z_{i,C}+z_{i,M})\bar{N} \quad \text{;}  \quad
	m_{i,t}^{*}=\frac{z_{i,M}}{z_{i,C}+z_{i,M}}wN_{i,t}  \quad  \text{;} \quad
	x^*_{i,j,t}=\frac{v_{i,j}}{p_j} \frac{z_{i,C}}{z_{i,C}+z_{i,M}}wN_{i,t} 
\label{optimal-values-consumption}
\end{equation}
As in the \gt, the underlying assumption is that money is comparable to other goods without any loss of generality as regarding, for instance, its decreasing marginal utility (Bentham 1748, Marshall 1890). Once determined the amount of money retained from consumption, each agent $i$ decides how to allocate it, as explained above, between liquid money, ($M^d_{2,i,t}$) and bonds, ($B_{i,t}$) in case either $L_{2 \, i,t}>r_t$ or $L_{2 \, i,t}<r_t$, respectively.

Aggregate values are the summation of individual ones. Thus, the aggregate consumption expenditure is $C_{t}=\sum_{i=0}^{n_p}c^*_{i,t}=\sum_{i=0}^{n_p}\sum_{j=0}^{n_c}p_j x^*_{i,j,t}$, the aggregate labour supply is $N^s_t=\sum_{i=0}^{n_p}N_{i,t}^{*} $, and the overall money demand is  $M^D_{t}=M^d_{1,t}+M^d_{2,t}$, where $M^d_{1,t}=B_{t}+C_{t}$ is the aggregate money demand for transactions, being $B_{t}=\sum_{i=0}^{n_p}B_{i,t}$ the aggregate bond purchases, and $M^d_{2,t}=\sum_{i=0}^{n_p}M^d_{2,i,t}$ is the aggregate speculative demand for money.  

\subsubsection{The Marginal Efficiency of Capital} \label{mek-section}
<<I define the marginal efficiency of capital as being equal to that rate of discount which would make te present value of the series of annuities given by the returns expected from the capital-asset during its life just equal to its supply price. This gives us the marginal efficiencies of particular capital-assets. The greatest of these marginal efficiencies can then be regarded as the marginal efficiency of capital in general>> (\gt, pp.$135$-$136$). As Keynes says, his definition of \textsc{mek} is close to the concept of 'marginal net efficiency' used by Marshall (in the sixth edition of his \textit{Principles of Economics}, dated 1910) and identical to the idea of 'rate of return over costs' used by Fisher (in his \textit{Theory of Interest}, dated 1930). The \textsc{mek} derives from animal spirits, thus distinguishing between uncertainty and risk (\gt, p.$161$): <<a spontaneous urge to action rather than inaction, and not as the outcome of a weighted average of quantitative benefits multiplied by quantitative probabilities>>. 

In order to compute it, each entrepreneur has to consider the \textit{prospective yields} obtainable from selling the output, deriving from the use of the capital asset (net of \textit{running expenses}) and its supply price, or replacement cost, which is <<the price which would just induce a manufacturer newly to produce an additional unit of such assets>> (\gt, p.$135$). Since the state of technology is exogenously given, the technology market is not represented in the \kcxx model. Each firm $h$ is endowed with an exogenously given supply price $\Psi^j_{h,t}(\tau_j)$, referred to the technology $j$ adopted by it, which will last for $\tau_j$ periods of time. For simplicity, it is assumed that adopted technologies imply that each k-firm uses natural resources and labour, whereas each c-firm uses all k-goods produced by k-firms and labour. In the \gt, the conditions of adoption of natural resources are not discussed; thus, in the \kcxx model, such aspects are neglected. Then, by following an approach similar to the one proposed by Minsky (1975), the \mekk of a firm $h$ adopting the technology $j$, i.e., $\text{\textsc{mek}}^{j}_{h,t}(\tau_j)$, is the rate which equals the value of the series of annuities, $Q_{h,t+1}, Q_{h,t+2}, ..., Q_{h,t+\tau_j}$ --which are <<prospective returns [...] from selling its output, after deducting the running expenses of obtaining that output>> (\gt, p.$135$)-- to the amount to be paid \textit{today} for the investment, $\Psi^j_{h,t}(\tau_j)$, and is computed as:
\begin{equation*} 
\Psi^j_{h,t}(\tau_j)=\sum_{s=t}^{t+\tau_j}\frac{Q_{h,s}}{\,\, \left[1+\text{\textsc{mek}}^{j}_{h,t}(\tau_j)\right]^s} 
\end{equation*}
Without loss of generality, assuming that technologies last for the same \textit{period} $\tau_j=\bar{\tau}$, i.e., the same <<time-units of notice of changes in the demand for it have to be given if it is to offer its maximum elasticity of employment>>  (\gt, p.$287$), and that each technology is used by a firm only, the previous equation can be simplified as:
\begin{equation} 
	\Psi_{h,t}=\sum_{s=t}^{t+\bar{\tau}}\frac{Q_{h,s}}{\,\, \left(1+\text{\textsc{mek}}_{h,t}\right)^s} 
	\label{mek-def}
\end{equation}
where, for simplicity, the indexation of firms and corresponding entrepreneurs will coincide; further, entrepreneurs of k-firms (c-firms) will be referred to as k-entrepreneurs (c-entrepreneurs). Eq.(\ref{mek-def}) requires, for each firm, the following series of ordered elements.
\begin{itemize}
	\item[a)] The amount of expected sales, i.e., the quantity to produce. Expectations have not been formally defined in the \gtt. In the \kcxx model, the amount of expected sales of each firm ($E_{t}[q_{h,t}]=\eta_h$) is an exogenous input, varied to compare the effects of the resulting \mekk over different scenarios.
	\item[b)] The optimal quantities of factors of production. Expected sales are used as the constraint for the optimization problem by which each firm minimizes the cost function dually associated with its technology. The standard Cobb-Douglas (\textsc{cd} henceforth) functional form has been used because of its tractability in calculations. All firms use labour ($L$); capital factors are, instead, differentiated: k-firms use natural resources ($R$), whereas c-firms use k-goods ($K$) produced by k-firms. Parameters of the \textsc{cd} technology are: the time index $\tau$; the shift parameter, $A_k$ ($A_c$), for the k-sector (c-sector); the exponents, i.e., $z_{h,R}$ and $z_{h,L}$  ($z_{h,K}$ and $z_{h,L}$), for k-firms (c-firms). It is assumed that $\sum z_{h,\bullet} < 1$, since Keynes often referred to the production functions in terms of decreasing returns (see, for instance, at pp.$17$, p.$92$, p.$122$, and p.$306$). Further, $z_{h,L}=\bar{z}_{L}$, $\forall h$. Realistically, firms adapt their production targets according to the availability of factors. Thus, each k-firm $h_k$ solves:
	\begin{equation*}
	\text{min}\quad  p_{\text{\textsc{r}}}R_{h_{k},t}+\bar{w}L_{h_{k},t} \quad \quad 
	\text{s.t.}\quad \quad A_k R^{z_{h,R}}_{h_{k},t}L^{z_{h,L}}_{h_{k},t} =E_{t}[q_{h_k,t}] 
\end{equation*}
	Natural resources management is not treated in the \gt; thus, $p_{\text{\textsc{r}}}$ is exogenously given. Optimal amounts of labour and natural resources for each k-firm are:
	\begin{align}
		L^*_{h_{k},t}=& \left[ \frac{1}{p_{\text{\textsc{r}}}} \right]^{\frac{z_{h,R}}{z_{h,R}+z_{h,L}}} \left[ \frac{E_{t}[q_{h_k,t+s}]}{A_k} \right]^{\frac{1}{z_{h,R}+z_{h,L}}} \left[ \frac{z_{h,L}}{\bar{w} \, z_{h,R}} \right]^{\frac{z_{h,R}}{z_{h,R}+z_{h,L}}}
		\label{optimal-labour-k-firms}\\
		R^*_{h_{k},t}=& \left[ \frac{1}{p_{\text{\textsc{r}}}} \right]^\frac{z_{h,L}}{z_{h,R}+z_{h,L}}  \left[ \frac{E_{t}[q_{h_k,t+s}]}{A_k} \right]^{\frac{1}{z_{h,R}+z_{h,L}}} \left[ \frac{\bar{w} \, z_{h,R}}{z_{h,L}}\right]^{\frac{z_{h,L}}{z_{h,R}+z_{h,L}}}
		\label{optimal-resources-k-firms}
	\end{align}
	Similarly, each c-firm $h_c$ optimizes by solving the following constrained minimization:
	\begin{equation*}
	\text{min} \quad \sum_{x_j \in G^k} p_j x_{j,h,t}+\bar{w}L_{h_{k},t} \quad \quad 
	\text{s.t.} \quad \quad A_c \left[ \prod_{x_j \in G^k}x^{ v_{h,j}}_{j,h,t}\right]^{z_{h,K}} \negthickspace \negthickspace \negthickspace L^{z_{h,L}}_{h_{k},t}  =E_{t}[q_{h_c,t}]  
\end{equation*}
	Optimal amount of labour, global capital, i.e., $K_{h_{k},t}=\prod_{x_j \in G^k}x^{ v_{h,j}}_{j,h,t}$, and optimal quantities of k-goods $(x^*_{h,j,t})$ for each c-firm are:
	\begin{align}
		L^*_{h_{c},t}=& \left[ \frac{1}{\prod_{j=0}^{n_k} \left(\frac{v_j}{p_j}\right)^{v_j}} \right]^{\frac{z_{h,K}}{z_{h,K}+z_{h,L}}} \left[ \frac{E_{t}[q_{h_c,t+s}]}{A_c} \right]^{\frac{1}{z_{h,K}+z_{h,L}}} \left[ \frac{z_{h,L}}{\bar{w} \, z_{h,K}} \right]^{\frac{z_{h,K}}{z_{h,K}+z_{h,L}}}
		\label{optimal-labour-c-firms} \\
		K^*_{h_{c},t}=& \left[\prod_{j=0}^{n_k} \left(\frac{v_j}{p_j}\right)^{v_j} \right]^\frac{z_{h,L}}{z_{h,K}+z_{h,L}}  \left[ \frac{E_{t}[q_{h_c,t+s}]}{A_c} \right]^{\frac{1}{z_{h,K}+z_{h,L}}} \left[ \frac{\bar{w} \, z_{h,K}}{z_{h,L}}\right]^{\frac{z_{h,L}}{z_{h,K}+z_{h,L}}}
		\label{optimal-capital-c-firms} \\
		x^*_{h,j,t}=& \frac{v_{h,j}}{p_j} \left[ \frac{E_{t}[q_{h_c,t+s}]}{A_c \, \Xi} \right]^{\frac{1}{z_{h,K}+z_{h,L}}}  \left[ \frac{\bar{w} \, z_{h,K}}{z_{h,L}}\right]^{\frac{z_{h,L}}{z_{h,K}+z_{h,L}}} \,\, \text{with} \,\, \Xi=\left[\prod_{j=0}^{n_k} \left(\frac{v_{h,j}}{p_j}\right)^{v_{h,j}} \right]^{z_{h,K}}
		\label{optimal-single-factor-c-firms}
	\end{align}
	where $v_{h,j}$ coefficients are defined as explained above and $\sum_{j} v_{h,j}=1$, $\forall h$ firms. 
	The labour force is then modelled as homogeneous and applied to decreasing returns technologies on hourly basis ($8$ hours per day), with $z_{i,\mathcal{L}}=\bar{z}_{\mathcal{L}}$ for all persons. <<We subsume, so to speak, the non-homogeneity of equally remunerated labour units in the equipment, which we regard as less and less adapted to employ the available labour units as output increases, instead of regarding the available labour units as less and less adapted to use a homogeneous capital equipment.>> (\gt, pp.$41$-$42$). 
	\item[c)] The cost functions. <<The amount paid out by the entrepreneur to the other factors of production in return for their services, which from their point of view is their income, we will call the factor cost of A>> (\gt, p.$53$). Whereas, <<an entrepreneur's user cost is by definition equal to $A_1+(G'-B')-G$, where $A_1$ is the amount of our entrepreneur's purchases from other entrepreneurs, $G$ the actual value of his capital equipment at the end of the period, and $G'$ the value it might have had at the end of the period if he had refrained from using it and had spent the optimum sum $B'$ on its maintenance and improvement>> (\gt, p.$66$). <<The sum of the factor cost $F$ and the user cost $U$ we shall call the prime cost of the output $A$>> (\gt, p.$53$). In the \kcx, dropping the depreciation, the prime cost of firm $h$, at time $t$, is defined as:
	\begin{equation}
		\textsc{pc}_{h,t}=\overbrace{W_{h,t}+\mathcal{R}_{h,t}}^{\textsc{fc}_{h,t}} +\overbrace{A_{1\, h,t}}^{\textsc{uc}_{h,t}}
	\label{prime-cost}
\end{equation}
where the first two addends are the factor cost $(\textsc{fc}_{h,t})$ and the last one is the user cost $(\textsc{uc}_{h,t})$. For k-firms, $A_{1\, h,t}=0$ and eq.(\ref{prime-cost}) reduces to $\textsc{pc}_{h,t}=W_{h,t}+\mathcal{R}_{h,t}$, where $\mathcal{R}_{h,t}=p_{\text{\textsc{r}}}R^*_{h_{k},t}$ is the cost of natural resources. For c-firms, since $R^*_{h,t}=0$, eq.(\ref{prime-cost}) becomes $\textsc{pc}_{h,t}=W_{h,t}+A_{1\, h,t}$, where $A_{1,\, h,t}=\sum_{x_j \in G^k} p_j x^*_{j,h,t}$ is the cost of k-goods purchases. Finally, for all firm types, $W_{h,t}$ stands for the set of expenditures required to remunerate the labour of both workers and the entrepreneur him/herself. This has been modelled by adding to the wage bill ($ \bar{w}L^*_{h,t}$) a personal variable of the entrepreneur, proportional to his/her optimism, measuring the desired remuneration for his/her own efforts as a multiple of the wage unit, computed as $(1+\mu_h)\bar{w}$, thus obtaining $W_{h,t}=[(1+\mu_h)+L^*_{h,t}]\bar{w}$. Keynes referred alto to \textit{supplementary} and \textit{windfall} costs, i.e., respectively, the involuntary excess of expected depreciation over the user cost (\gt, p.$56$) and the unforeseen change in the value of the equipment (\gt, p.$57$). Without loss of generality, they will be neglected.
	\item[d)] Pricing. The <<aggregate supply function for a given firm (and similarly for a given industry or for industry as a whole) is given by $Z_r = \phi_r(N_r)$ where $Z_r$ is the proceeds (net of user cost) the expectation of which will induce a level of employment $N_r$. If, therefore, the relation between employment and output is such that an employment $N_r$ results in an output $O_r$ where $O_r = \psi_r(N_r)$, it follows that $p=\frac{Z_r+U_r(N_r)}{O_r}=\frac{\phi_r(N_r)+U_r(N_r)}{\psi_r(N_r)}$ is the ordinary supply curve, where $U_r(N_r)$ is the (expected) user cost corresponding to a level of employment $N_r$.>> (\gt, p.$44$). The pricing of firm $h$ is:
	\begin{equation}
		p_{h,t}=\frac{\textsc{pc}_{h,t}}{E_t[q_{h,t}]} 
		\label{pricing}
	\end{equation} 
	\item[e)] The amount of expected revenues ($E_t[\mathcal{S}_{h,t}]$) is computed by each firm $h$ simply as: 
	\begin{equation}
		E_t[\mathcal{S}_{h,t}]=p_{h,t}E[q_{h,t}]
	\end{equation}
	\item[f)] Proceeds. Each firm $h$ computes the series of prospective returns for each of $\bar{\tau}$ time steps, i.e., $Q_{h,t+1}, Q_{h,t+2}, ..., Q_{h,t+\bar{\tau}}$, as 
	\begin{equation}
		Q_{h,t}=E_t[\mathcal{S}_{h,t}]-\textsc{pc}_{h,t}
		\label{eq:proceeds}
	\end{equation} 
	\item[g)] The \mek. Finally, each firm can finally obtain its own $\mek_{h,t}$ by computing eq.(\ref{mek-def}), in which above-defined elements, eqs.(\ref{optimal-labour-k-firms})-(\ref{eq:proceeds}), can be used as input. 
\end{itemize}
Each entrepreneur can finally decide whether to invest or not according with the fact that the marginal efficiency of capital related to the investment for his/her firm $h$ is greater than both his/her personal preference to liquidity and the interest rate, being the alternatives either a financial investment (in bonds, yielding the interest rate), or to held liquidity for speculative reasons. The decision rule adopted by the entrepreneur $h$, owner of firm $h$, is:
\begin{equation}
	\begin{array}{lccl}
		\text{\textsc{if}} \,\,\,\,& \text{\textsc{mek}}_{h,t} > \, \text{max} \, [L_{2 \, h,t}(r_t),r_t]& \rightarrow &  I_{h,t} \\
		\text{\textsc{if}} \,\,\,\,& r_t > \, \text{max} \, [L_{2 \, h,t}(r_t), \text{\textsc{mek}}_{h,t}] & \rightarrow &  B_{h,t}\\
		\text{\textsc{if}} \,\,\,\,& L_{2 \, h,t}(r_t) > \, \text{max} \, [\text{\textsc{mek}}_{h,t} ,r_t] & \rightarrow &  M^d_{2,h,t}
	\end{array}
	\label{investment-choice}
\end{equation}
In case the investment is deliberated, actual quantities are computed by using same equations adopted to cast expected values. The aggregate investment expenditure is obtained by summing individual investments of entrepreneurs, $I_t=\sum_{h=1}^{n_g}I_{h,t}$. Aggregate investment is recorded as the sum of technological investment and the value variation of inventories. Thus, <<it is obvious that the actual rate of current investment will be pushed to the point where there is no longer any class of capital-asset of which the marginal efficiency exceeds the current rate of interest>> (\gt, p.$136$). Expectedly, for any given level of $\textsc{mek}_{t}$, $ I_{t}$ results to be inversely correlated to the interest rate.

\subsection{Dependent variables}
<<Our dependent variables are the volume of employment and the national income (or national dividend) measured in wage-units.>> (\gt, p.$245$). The ``volume of employment'' is the number of workers actually employed in any of existing firms, i.e., $ N_t= \sum_{h=1}^{g_n}N_{h,t}$, being $N_{h,t}$ the count of personnel hired by firm $h$. Since the unemployment is $U_t=n_p-N_t$, the unemployment rate is computed as: $u_t=U_t /n_p=1-(N_t/n_p)$. Similarly, the national income is obtained by considering the summation of all distributed income in the economy, for all persons, i.e. $Y_t=\sum_{i=0}^{n_p}y_{i,t}$, where $y_{i,t}$ represents wages for workers and profits for entrepreneurs, plus the non-distributed profits of income-relevant firms (i.e., wages, owner rents, and retained firms' profits generated by current production and sales). The accounting block of the model records aggregate expenditure as
\[
Z_t=C_t+I_t
\]
where
\(I_t=I^{tech}_t+\Delta Inv_t\) combines technological investment and inventory investment. Aggregate saving includes household saving and the non-distributed profits of income-relevant firms. Hence the equality
\[
S_t=I_t
\]
is an ex-post accounting result of the simulated sequence, not an equilibrium condition imposed by the interest rate. The corresponding investment multiplier is computed as
\[
\kappa_t=\frac{1}{1-C_t/Y_t}
\]
and, whenever \(Y_t=Z_t\), the model verifies \(Y_t=\kappa_t I_t\) up to numerical rounding.

\section{Results}\label{sec:results}

The simulations reported in this section are static one-period experiments. The model is initialized, firms form sales expectations, compute the corresponding optimal input requirements, evaluate expected costs and proceeds, and then compare their marginal efficiency of capital with the relevant financial threshold. Production takes place only when the expected return is not smaller than the maximum between the interest rate and the entrepreneur's liquidity preference. The relevant threshold is therefore
\[
h=\max\{r,L_2\}
\]
The sequence is important. k-firms move first and determine the availability of capital goods. c-firms then form their production plans subject to the availability of these k-goods. Households allocate their income between consumption, speculative money demand and bonds only after income has been generated. Hence, the model is designed to make the Keynesian chain running from expectations to investment, employment, income and monetary allocation explicit.

All monetary values in the tables and figures are expressed in wage units. In the figures, $Y^w$, $C^w$, $S^w$ and $I^w$ are reported in millions of wage units. Monte Carlo results are averages over the replications indicated in the appendix. The final simulation batches are fully balanced: the calibration refinement contains 56 output files, the sectoral animal-spirits experiment contains 70 output files, and the $r\times L_2$ maps contain 280 output files. The latter are explicitly separated into a homogeneous benchmark and three heterogeneous regimes: animal spirits, liquidity preference and technology prices.

\begin{table}[t]\scriptsize
\centering
\begin{tabular}{lll}
\toprule
Variable & Description & Value \\
\midrule
$n_p$ & persons & 5,000 \\
$n_k$ & k-goods and k-firms & 5 \\
$n_c$ & c-goods and c-firms & 15 \\
$AnSp_k=AnSp_c$ & baseline animal spirits & 5 \\
$I^c_k$ & technology price of k-firms & 30,000 \\
$I^c_c$ & technology price of c-firms & 570,000 \\
$\overline{\mathrm{MEK}}$ & MEK upper bound & 5.5\% \\
$\underline{\mathrm{MEK}}$ & MEK lower bound & 0.5\% \\
$\bar{w}$ & money wage & 10 \\
$\bar{\tau}$ & MEK horizon & 5 periods \\
\bottomrule
\end{tabular}
\caption{Baseline calibration used for the final static simulations.}
\label{tab:kcx-calibration-final}
\end{table}

\subsection{Calibration and simulation design}

The final calibration is reported in Table~\ref{tab:kcx-calibration-final}. The calibration of the technological supply prices was selected after an extensive exploration of the parameter space, varying animal spirits, technological supply prices, the interest rate and liquidity preference. The objective was to place the model in a non-degenerate region, where the two sectors remain economically connected and the financial threshold is able to affect investment decisions. If technological supply prices are too low, firms' marginal efficiencies of capital remain at the imposed ceiling and the threshold cannot select among them. If they are too high, the c-sector may disappear too easily, either because investment becomes financially unprofitable or because the sequential dependence on the k-sector generates binding bottlenecks. The selected calibration therefore allows the model to be tested in a region where both sectors remain operative, while firms can still be progressively filtered as \(h=\max\{r,L_2\}\) increases.

The lower bound of the marginal efficiency of capital is $0.5\%$ and the upper bound is $5.5\%$. This ensures that the range of $r$ and $L_2$ used in the simulations can intersect the distribution of investment returns. The model therefore allows the threshold $h$ to generate either a binary transition, when agents are homogeneous, or a gradual selection, when technology prices or liquidity preferences are heterogeneous.

\subsection{Accounting identities and multipliers}

Before discussing the behavioural mechanisms, it is useful to record the accounting structure generated by the sequential simulation. In the no-government specification used here, aggregate expenditure is
\[
Z_t=C_t+I_t
\]
where investment includes both technological expenditure and the value variation of inventories,
\[
I_t=I^{tech}_t+\Delta Inv_t
\]
The model computes saving after income generation, consumption, firms' sales, inventory accumulation and retained profits have been recorded. Consequently,
\[
S_t=I_t
\]
holds ex post as an accounting result, not as an equilibrium condition imposed through the interest rate. This is the Keynesian reversal of the classical interpretation: saving and investment are determinates of the system, while income is the magnitude through which their equality is realised. This is also the point emphasised by Leontief's discussion of Keynes's rejection of the homogeneity postulate.

\begin{figure}[t]
	\centering
	\includegraphics[width=.45\textwidth]{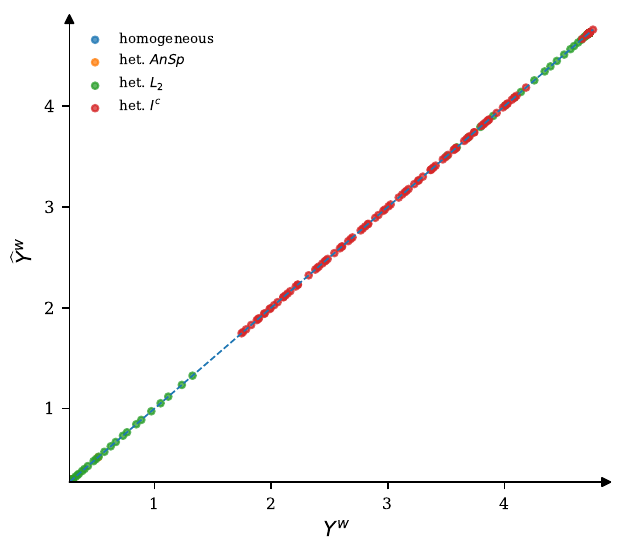}
	\caption{Investment multiplier verification. The vertical axis reports $\widehat Y^w=I^w/(1-C^w/Y^w)$; the horizontal axis reports simulated income $Y^w$.}
	\label{fig:kcx-income-multiplier}
\end{figure}

The same accounting closure recovers the investment multiplier. Defining
\[
c_t=\frac{C_t}{Y_t}
\]
the implied multiplier is
\[
\kappa_Y=\frac{1}{1-c_t}
\]
and the simulated accounts satisfy
\[
\widehat Y_t=\frac{I_t}{1-C_t/Y_t}=Y_t
\]
up to numerical rounding. Hence, the multiplier is not inserted as an external formula; it is recovered from the simulated values of consumption, investment and income. Table~\ref{tab:kcx-accounting-multiplier} reports this verification for representative regimes, and Figure~\ref{fig:kcx-income-multiplier} shows that the simulated income and multiplier-implied income lie on the same line.

\begin{table}[t]\scriptsize
\centering
\begin{tabular}{lrrrrrrr}
\toprule
Scenario & $Y^w$ & $C^w$ & $I^w$ & $S^w-I^w$ & $C^w/Y^w$ & $\kappa$ & $Y^{w,*}-Y^w$ \\
\midrule
Homogeneous, $h=0.5\%$ & 4.726 & 1.792 & 2.934 & -0.00667 & 0.379 & 1.611 & 0.0107 \\
Homogeneous, $h=3.8\%$ & 0.293 & 0.000 & 0.293 & 0 & 0.000 & 1.000 & 0 \\
Het. $I^c$, $h=0.5\%$ & 4.759 & 1.792 & 2.968 & -0.00667 & 0.377 & 1.604 & 0.0107 \\
Het. $I^c$, $h=1\%$ & 3.934 & 1.500 & 2.434 & -0.000239 & 0.381 & 1.616 & 0.000387 \\
Het. $I^c$, $h=3\%$ & 2.892 & 0.988 & 1.904 & 1.72e-08 & 0.334 & 1.519 & -3.59e-08 \\
Het. $I^c$, $h=5\%$ & 1.746 & 0.363 & 1.383 & 2.56e-08 & 0.191 & 1.262 & -7.45e-09 \\
Het. $L_2$, $h=3.8\%$ & 1.326 & 0.335 & 0.991 & 1.48e-08 & 0.134 & 1.338 & -1.33e-08 \\
\bottomrule
\end{tabular}
\caption{Accounting consistency and investment multiplier verification. Monetary magnitudes are reported in millions of wage units except residuals.}
\label{tab:kcx-accounting-multiplier}
\end{table}

\begin{figure}[t]
	\centering
	\includegraphics[width=.85\textwidth]{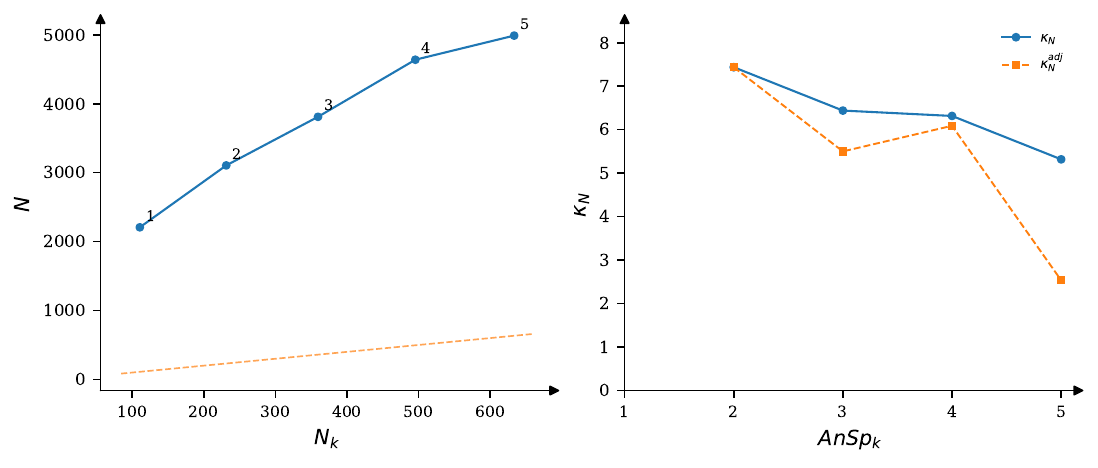}
	\caption{Kahn-type employment multiplier. Panel A relates total employment to primary employment in the k-sector; Panel B reports cumulative and adjacent employment multipliers.}
	\label{fig:kcx-kahn-multiplier}
\end{figure}

The sectoral structure of the model also makes it possible to compute a Kahn-type employment multiplier. Since k-firms are the investment-goods sector, their employment is treated as primary employment. Along the section $AnSp_c=5$, the cumulative employment multiplier is defined as
\[
\kappa_N(\eta_k)=
\frac{N(\eta_k,5)-N(1,5)}{N_k(\eta_k,5)-N_k(1,5)}
\]
for $\eta_k>1$. This measures the increase in total employment associated with an increase in primary employment in the investment-goods sector. Table~\ref{tab:kcx-kahn-employment-multiplier} and Figure~\ref{fig:kcx-kahn-multiplier} show that the multiplier is larger than one throughout the relevant range, and declines only as the system approaches full employment.

\begin{table}[t]\scriptsize
\centering
\begin{tabular}{rrrrr}
\toprule
$AnSp_k=AnSp_c$ & $\Delta N_k$ & $\Delta N_c$ & $\Delta N$ & $\kappa_N^{\Delta}$ \\
\midrule
2 & 120.6 & 823.2 & 943.8 & 7.83 \\
3 & 250.2 & 1699.0 & 1949.2 & 7.79 \\
4 & 385.3 & 2594.6 & 2979.9 & 7.73 \\
5 & 523.0 & 3510.2 & 4033.2 & 7.71 \\
\bottomrule
\end{tabular}
\caption{Employment multiplier in the Kahnian sense. Increments are computed relative to the low-expectation benchmark $AnSp_k=AnSp_c=1$, using the Monte Carlo animal-spirits matrix.}
\label{tab:kcx-kahn-multiplier}
\end{table}

\subsection{Animal spirits and sectoral complementarity}

\begin{figure}[t]
	\centering
	\includegraphics[width=.95\textwidth]{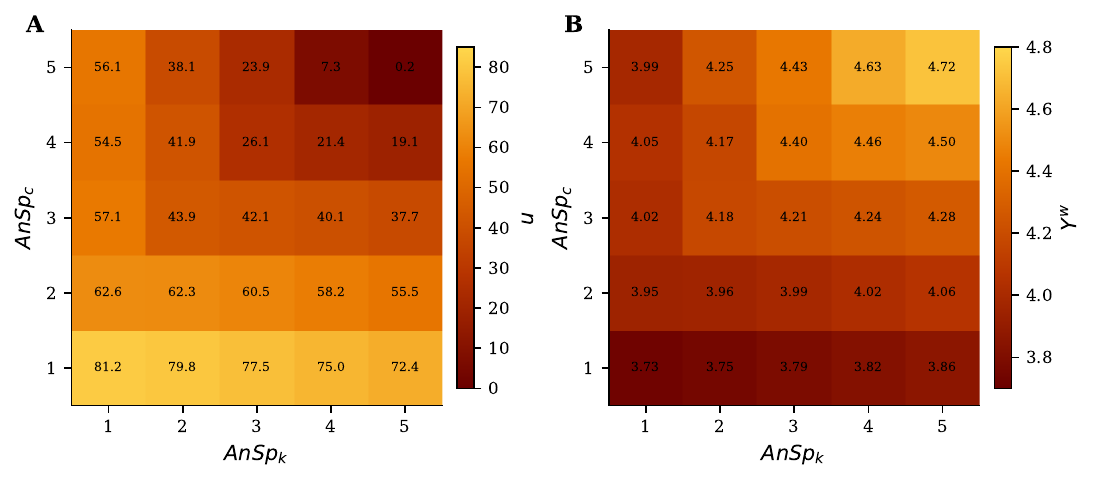}
	\caption{Sectoral complementarity in the animal-spirits matrix. Panel A reports the unemployment rate $u$; Panel B reports income $Y^w$.}
	\label{fig:kcx-sector-complementarity}
\end{figure}

The first result concerns the role of entrepreneurs' animal spirits. The experiment varies $AnSp_k$ and $AnSp_c$ over the grid $\{1,2,3,4,5\}\times\{1,2,3,4,5\}$. Figure~\ref{fig:kcx-sector-complementarity} reports the unemployment rate and aggregate income. The relationship is monotone in both directions: higher animal spirits raise income and reduce unemployment. More importantly, the response is complementary. High expectations in the c-sector are insufficient when the k-sector is pessimistic and does not produce enough capital goods. Similarly, a strong k-sector alone cannot generate full employment if expected sales in the c-sector remain weak.
\begin{table}[t]\scriptsize
\centering
\begin{tabular}{rrrrrr}
\toprule
$AnSp_k$ & $AnSp_c$ & $Y^w$ & $N$ & $u$ & $Y_c/Y_c^{e,0}$ \\
\midrule
$1$ & $1$ & $3.729$ & $954.6$ & $81.23\%$ & $1.004$ \\
$1$ & $5$ & $3.989$ & $2206.7$ & $56.09\%$ & $0.425$ \\
$5$ & $1$ & $3.861$ & $1396.2$ & $72.37\%$ & $1.002$ \\
$5$ & $5$ & $4.723$ & $4987.8$ & $0.24\%$ & $0.997$ \\
\bottomrule
\end{tabular}
\caption{Key Monte Carlo cells in the sectoral animal-spirits matrix. $Y^w$ is reported in millions of wage units.}
\label{tab:kcx-key-ansp-cells}
\end{table}

\begin{figure}[t]
	\centering
	\includegraphics[width=.6\textwidth]{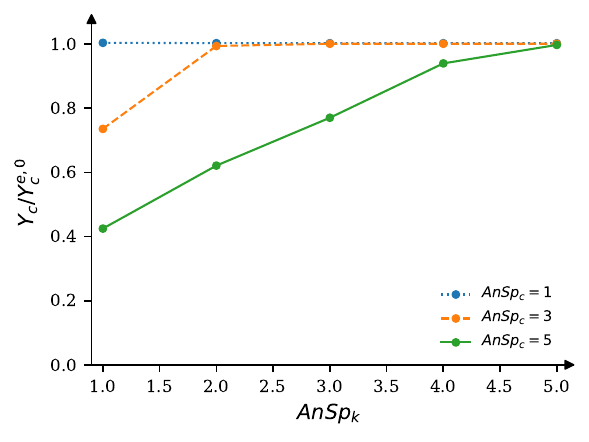}
	\caption{Realization of expected c-output for different levels of k-sector animal spirits.}
	\label{fig:kcx-c-realization}
\end{figure}

Table~\ref{tab:kcx-key-ansp-cells} isolates four cells of the same matrix. Moving from $(AnSp_k,AnSp_c)=(1,1)$ to $(5,5)$ reduces unemployment from more than $81\%$ to almost zero. Yet the asymmetric cells show that optimism concentrated in one sector alone does not reproduce the full-employment outcome. This is a structural implication of the sequential architecture of the model.
Figure~\ref{fig:kcx-c-realization} shows the same mechanism from the side of realized c-output. When $AnSp_c=5$, the realized-to-autonomous c-output ratio rises from approximately $0.425$ at $AnSp_k=1$ to almost one at $AnSp_k=5$. The interpretation is direct: even when c-firms expect high sales, their plans cannot be realized if k-goods are not sufficiently available.

\subsection{The financial threshold and the \textsc{mek}}

The second result concerns the financial condition for investment. Since firms compare the marginal efficiency of capital with both the interest rate and liquidity preference, the operative threshold is $h=\max\{r,L_2\}$. Table~\ref{tab:kcx-threshold-main} summarizes the four regimes used to isolate different sources of heterogeneity.

\begin{table}[t]\scriptsize
\centering
\begin{tabular}{lrrrrrr}
\toprule
 & \multicolumn{6}{c}{$h=\max(r,L_2)$} \\
\cmidrule(lr){2-7}
Regime & 0.5\% & 1\% & 2\% & 3\% & 4\% & 5\% \\
\midrule
\multicolumn{7}{l}{\emph{Active c-firms}} \\
Homogeneous & 15.00 & 15.00 & 15.00 & 15.00 & 0.00 & 0.00 \\
Het. $AnSp$ & 15.00 & 15.00 & 15.00 & 15.00 & 0.00 & 0.00 \\
Het. $L_2$ & 15.00 & 15.00 & 15.00 & 14.68 & 2.30 & 0.06 \\
Het. $I^c$ & 14.82 & 12.55 & 10.76 & 9.16 & 6.93 & 5.28 \\
\addlinespace[0.3em]
\multicolumn{7}{l}{\emph{Unemployment rate}} \\
Homogeneous & 0.00\% & 0.00\% & 0.00\% & 0.00\% & 87.25\% & 87.25\% \\
Het. $AnSp$ & 0.26\% & 0.29\% & 0.26\% & 0.27\% & 87.27\% & 87.25\% \\
Het. $L_2$ & 0.00\% & 0.00\% & 0.00\% & 1.82\% & 73.87\% & 88.41\% \\
Het. $I^c$ & 1.04\% & 14.18\% & 24.59\% & 33.89\% & 46.92\% & 56.52\% \\
\bottomrule
\end{tabular}
\caption{Financial threshold, active c-firms and unemployment. Values are averages over the relevant grid cells.}
\label{tab:kcx-threshold-main}
\end{table}

\begin{figure}[t]
	\centering
	\includegraphics[width=.95\textwidth]{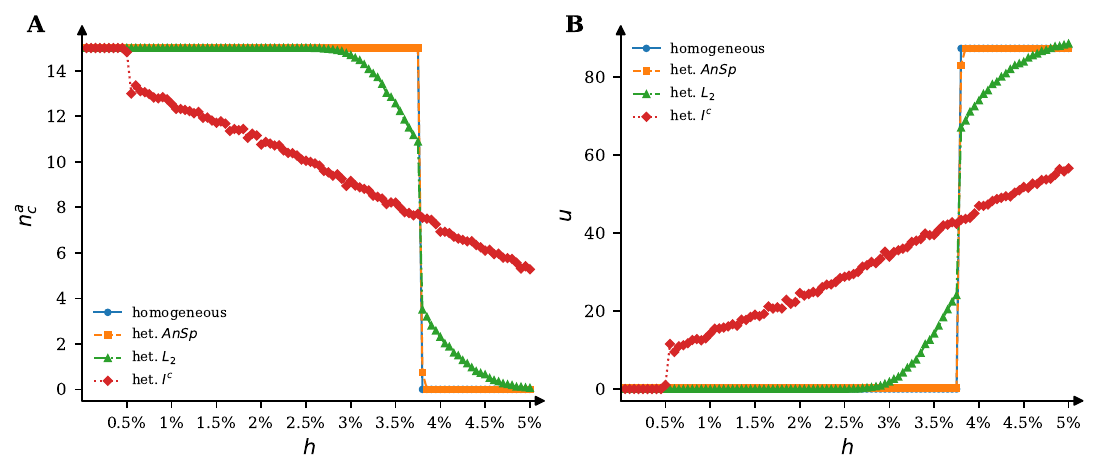}
	\caption{Homogeneous and heterogeneous threshold regimes. Panel A reports active c-firms; Panel B reports the unemployment rate.}
	\label{fig:kcx-threshold-regime-comparison}
\end{figure}

Figure~\ref{fig:kcx-threshold-regime-comparison} makes the distinction between homogeneous and heterogeneous settings explicit. The homogeneous economy is the limiting benchmark: with identical firms and identical thresholds, the investment decision changes almost simultaneously across the c-sector. Heterogeneity in animal spirits mainly affects the real scale of expected production; it does not by itself produce a wide dispersion of financial returns. Heterogeneity in liquidity preference smooths the threshold because entrepreneurs face different individual values of $L_2$. Heterogeneity in the technology price $I^c$ produces the clearest gradual MEK selection, since firms face different investment prices for comparable expected proceeds.

In the homogeneous benchmark, the transition is nearly discontinuous. When $h$ is below the common c-firm marginal efficiency, all c-firms are active and unemployment is zero. Once the threshold reaches the relevant value, the c-sector stops and unemployment jumps to about $87\%$. This is the clean theoretical case: identical firms facing the same threshold make the same investment decision.

\begin{figure}[htbp]
	\centering
	\begin{minipage}[b]{0.45\textwidth}
		\centering
		\includegraphics[width=\textwidth]{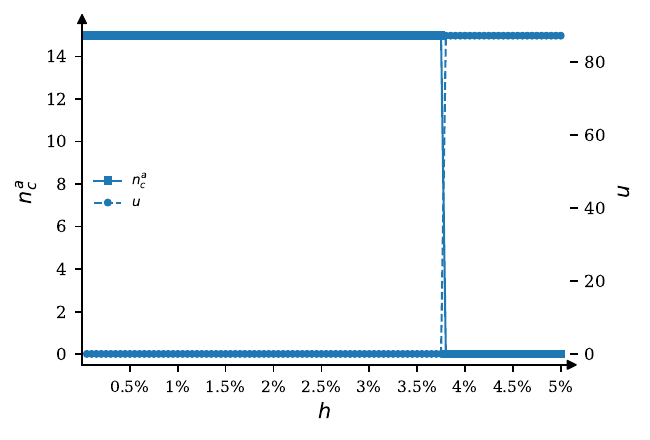}
		\caption{Homogeneous benchmark: common threshold and c-firm exit.}
		\label{fig:kcx-homogeneous-threshold}
	\end{minipage}
	\begin{minipage}[b]{0.45\textwidth}
		\centering
		\includegraphics[width=\textwidth]{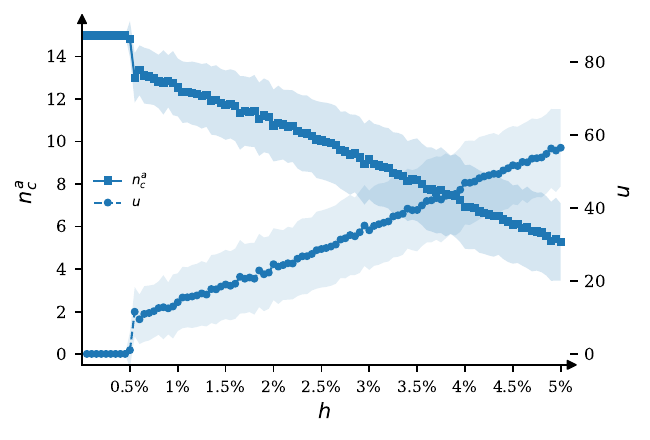}
		\caption{MEK selection with heterogeneous technology prices.}
		\label{fig:kcx-ic-selection}
	\end{minipage}
\end{figure}

The heterogeneous investment-cost case is the central result for the marginal efficiency of capital. When $I^c$ differs across firms, the same threshold no longer selects all firms simultaneously. Instead, c-firms are filtered progressively. Figure~\ref{fig:kcx-ic-selection} shows that the number of active c-firms declines smoothly as $h$ rises, while unemployment increases. This is the mechanism that translates a higher interest rate or a higher liquidity-preference threshold into lower investment, income and employment.

\begin{figure}[t]
	\centering
	\includegraphics[width=.6\textwidth]{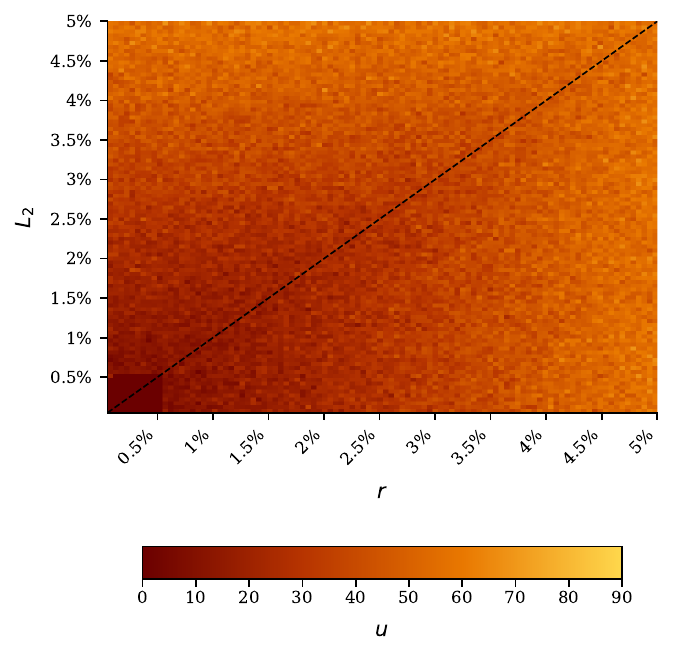}
	\caption{Unemployment in the \(r\times L_2\) plane with heterogeneous technology prices.}
	\label{fig:kcx-ic-u-map}
\end{figure}

The same result can be represented over the full monetary plane. Figure~\ref{fig:kcx-ic-u-map} shows unemployment in the \(r\times L_2\) space when technology prices are heterogeneous. The relevant boundary is governed by \(h=\max\{r,L_2\}\): unemployment rises whenever either the interest rate or liquidity preference pushes the financial threshold above the marginal efficiency of a growing fraction of c-firms.

Heterogeneity in animal spirits, by itself, does not generate the same smooth financial selection. It works mainly through the real channel, namely expected output and factor demand. Heterogeneity in liquidity preference instead smooths the threshold because entrepreneurs differ in the value of $L_2$ entering the investment decision. These two cases are reported in the appendix and serve as robustness checks around the central heterogeneous-$I^c$ mechanism.

\subsection{Liquidity preference and the monetary allocation}

The third result concerns households' monetary allocation. Once income has been generated, each household allocates its residual liquidity between speculative money demand and bonds. The model therefore distinguishes the productive investment threshold from the portfolio allocation rule. In the homogeneous case, the rule is sharp: if $L_2>r$, residual liquidity is allocated to speculative money demand; if $L_2<r$, it is allocated to bonds; if $L_2=r$, it is split between the two.

\begin{figure}[t]
	\centering
	\includegraphics[width=.6\textwidth]{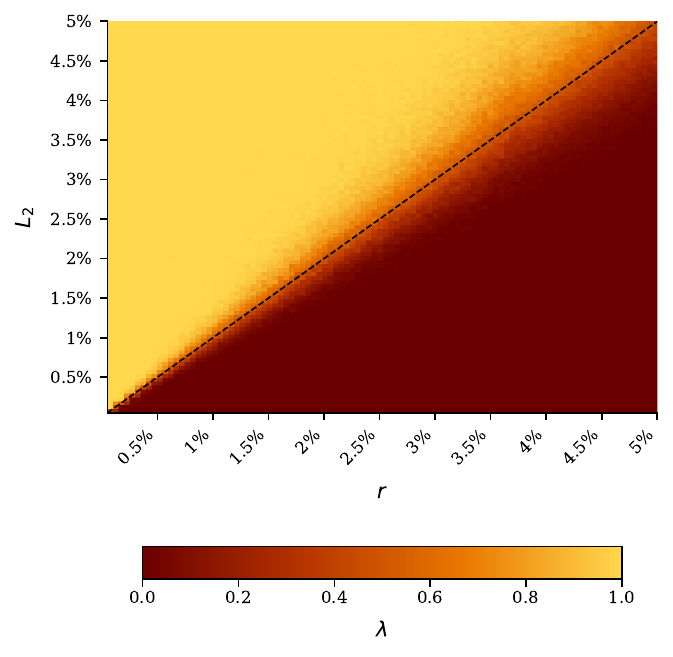}
	\caption{Liquidity-preference regimes in the $r\times L_2$ plane. The color reports $\lambda=M^d_2/(M^d_2+B)$.}
	\label{fig:kcx-liquidity-allocation}
\end{figure}

Figure~\ref{fig:kcx-liquidity-allocation} reports the index
\[
\lambda=\frac{M^d_2}{M^d_2+B}
\]
in the $r\times L_2$ plane. The diagonal $r=L_2$ separates the two monetary regimes. Values above the diagonal correspond to a prevalence of speculative money demand; values below the diagonal correspond to bond purchases. With heterogeneous liquidity preferences, the transition becomes a band rather than a knife-edge, but the same rule remains visible.

\begin{figure}[t]
	\centering
	\includegraphics[width=.6\textwidth]{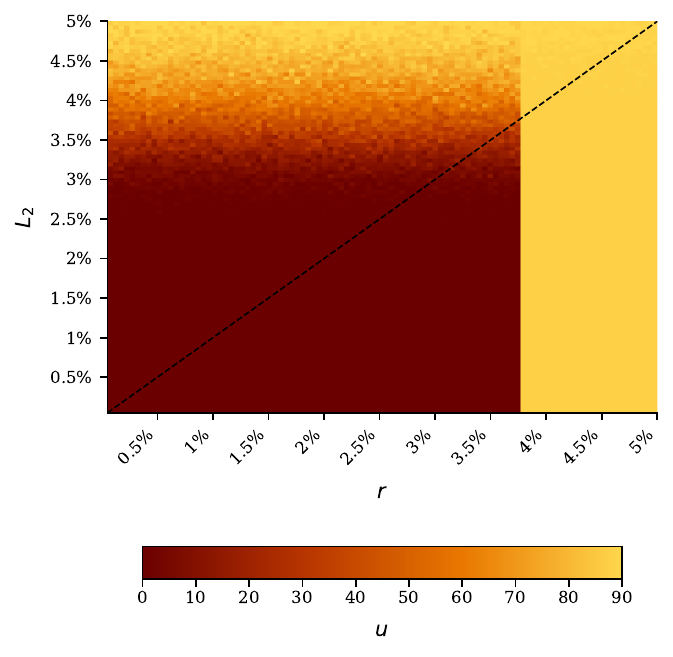}
	\caption{Unemployment in the \(r\times L_2\) plane with heterogeneous liquidity preferences.}
	\label{fig:kcx-lp-u-map}
\end{figure}

Liquidity preference also affects production through the investment threshold. Figure~\ref{fig:kcx-lp-u-map} reports unemployment in the same \(r\times L_2\) plane when liquidity preferences are heterogeneous. The result complements Figure~\ref{fig:kcx-liquidity-allocation}: the difference \(L_2-r\) governs the allocation of residual liquidity, while the level of \(\max\{r,L_2\}\) governs the financial feasibility of investment.

\section{Conclusions}

This article has developed a formal and computational reconstruction of Keynes's \gt, implemented in the \kcxx model. The aim has been to preserve the sequential structure of  Keynesian theory. All agents, both households and firms, first resolve their portfolio allocation between consumption, liquidity, bonds, and real investment, and only then, on the basis of those monetary decisions, do entrepreneurs acquire factors, produce, and distribute incomes, which recipients in turn allocate again. The order of decisions is therefore not a technical detail. It is the mechanism through which monetary variables become real determinants of income and employment.

The simulations support three main conclusions. First, liquidity preference governs the allocation of income between real activities, speculative money balances and bonds, and in doing so determines the monetary environment that influences expenditure elements. This allocation is the primary decision, because it establishes the minimum yield that entrepreneurs require before committing resources to production. In heterogeneous settings this relation becomes a smooth aggregate frontier rather than a sharp individual switch, so that the same mechanism generates a gradual and continuous distribution of investment decisions across the population of firms. The model thus reproduces the Keynesian claim that money is not neutral: monetary expectations shape both investment decisions and the composition of demand, and the interest rate is a monetary phenomenon that precedes and conditions the real one.

Second, animal spirits affect aggregate outcomes through a genuinely real channel. Higher expected sales increase desired production, factor demand, income, and employment. The sectoral structure is also essential: the consumption sector cannot realise its own expectations unless the capital-goods sector has previously generated the required productive inputs. The model therefore shows that aggregate employment depends not only on the level of optimism, but also on the coordination of expectations across sectors.

Third, the financial threshold above which investment becomes profitable depends jointly on the interest rate and on agent-specific liquidity preference, with the binding constraint given by whichever of the two is higher. When the marginal efficiency of capital falls below this threshold, investment is not undertaken. In the homogeneous case, this produces an almost discontinuous transition between full engagement and complete withdrawal from production. In heterogeneous settings, especially when the price of technology differs across firms, the same mechanism becomes gradual: firms are progressively selected out of production as the threshold rises. Employment is therefore not determined by a real full-employment benchmark; it depends on whether expected yields are sufficiently high relative to the monetary return required by agents, which is itself shaped by the state of liquidity preference in the economy.

The central result is therefore the following. In a sequential economy, monetary variables do not merely determine nominal magnitudes: they enter the causal structure of production. Expectations determine the scale of desired activity, but the interest rate and liquidity preference determine whether the expected yields are sufficient for investment to occur. Once investment decisions are filtered through this monetary threshold, aggregate income and employment are determined at or below full employment, depending on the configuration of expectations, financial conditions, and heterogeneity. The \kcxx model thus shows that sequentiality is the structural feature that distinguishes Keynesian from equilibrium macroeconomics, i.e., the very feature that this paper set out to formalise from the opening pages of the \gt. 

The model also clarifies why heterogeneity matters. Homogeneous agents generate clear benchmark cases, useful for isolating Keynesian mechanisms in their purest form. Heterogeneous agents, however, transform those mechanisms into empirically more plausible aggregate patterns: discontinuous thresholds become gradual transitions, and individual monetary choices become aggregate distributions. The \kcxx model thus provides a micro-designed representation of the \gt in which macroeconomic outcomes emerge from decentralised, sequential, and monetary decisions.
\newpage

\section*{Acknowledgments}
\noindent The quotations cited in the paper are from \textit{The Royal Economic Society \copyright$\,$1973 Edition}, printed by The MacMillan Press, Ltd.
The author is grateful to Domenico Delli Gatti, Mauro Gallegati, Marcello Signorelli and Roberto Zanola for their comments on earlier versions of the manuscript. Any errors or limitations in the paper and model are solely the responsibility of the author.

\section*{Funding}
\noindent The Author acknowledges the support of PTR and Piaceri Funding schemes provided by the University of Catania.

\section*{Conflict of Interests} 
\noindent The Author declares no conflict of interests/competing interests.

\newpage

\section*{References}

Akerlof, G.A. (2002).  Behavioral macroeconomics and macroeconomic behavior. \textit{American Economic Review}, vol.92(3), pp.411-- 433. \\

Alexander, S.S. (1940).  Mr. Keynes and Mr. Marx. \textit{The Review of Economic Studies}, vol.7(2), pp.123--135. \\

Ashraf, Q., Gershman, B. and Howitt, P. (2016). How inflation affects macroeconomic performance: An agent-based computational investigation. \textit{Macroeconomic dynamics}, vol.20(2), pp.558--581. \\

Ashraf, Q., Gershman, B. and Howitt, P. (2017). Banks, market organization, and macroeconomic performance: an Agent-based computational analysis. \textit{Journal of Economic Behavior \& Organization}, vol.135, pp.143--180. \\

Axtell, R.L. (2007). What economic agents do: How cognition and interaction lead to emergence and complexity. \textit{The Review of Austrian Economics}, vol.20, pp.105--122.\\

Blaug, M. (1974). \textit{The Cambridge Revolution: Success or Failure?} Institute of Economic Affairs. London.\\

Blaug, M. (1980). \textit{The Methodology of Economics}. Cambridge University Press. Cambridge.\\

Blinder, A. S. (1987). Keynes, Lucas, and scientific progress. \textit{The American Economic Review}, vol.77(2), pp.130--136. \\

Biondo A.E. (2023). Mr.Keynes and the... Complexity! A suggested agent-based version of the General Theory of Employment, Interest and Money. 	arXiv:2303.00889 [econ.TH]. https://doi.org/10.48550/arXiv.2303.00889\\

Buchanan, J. M., Burton, J., and Wagner, R. E. (1978). \textit{The Consequences of Mr Keynes}. London: Institute of Economic Affairs.\\

Carabelli, A. (1988). \textit{On Keynes's Method}. Palgrave Macmillan, NY.\\

Chick, V. (1983). \textit{Macroeconomics after Keynes: a reconsideration of the general theory}, \textsc{mit} Press. \\

Christiano, L.J., Eichenbaum, M.S., and Trabandt M. (2018). On DSGE Models. \textit{Journal of Economic Perspectives}, vol. 32 (3), pp.113--140\\

Cincotti, S.,  Raberto, M. and Teglio, A. (2010). Credit money and macroeconomic instability in the agent-based model and simulator \textsc{eurace}. \textit{Economics}, vol.4(1), pp.1--32. \\

Cincotti, S.,  Raberto, M. and Teglio, A.  (2012). The \textsc{eurace} macroeconomic model and simulator. In Agent-based Dynamics, Norms, and Corporate Governance. \textit{The proceedings of the16-th World Congress of the International Economic Association, Palgrave}, vol.2. \\

Cohen, A. and Harcourt, G. Whatever Happened to the Cambridge Capital Controversies? \textit{Journal of Economic Perspectives}, Winter 2003, pp.199--214.\\

Colander, D. (2010). The Keynesian Method, Complexity, and the Training of Economists. \textit{Middlebury College Economics Discussion Paper} NO. 10-35.\\

Crotty, J.R. (1983). On Keynes and capital flight. \textit{Journal of Economic Literature}, vol.21(1), pp.59--65. \\

Cutler, T., Williams, J. and Williams, K. (2013). \textit{Keynes, Beveridge and beyond},  Routledge, London. \\

Davidson, P. (2007). \textit{John Maynard Keynes}, Palgrave MacMillan, NY.\\

Dawid, H. and Delli Gatti, D. (2018). Agent-based macroeconomics. \textit{Handbook of computational economics}, vol.4, pp.63--156. \\

Dawid, H., Harting, P. and Neugart, M. (2018). Cohesion Policy and Inequality Dynamics: Insights from a Heterogeneous Agents Macroeconomic Model. \textit{Journal of Economic Behavior \& Organization}, vol.150, pp. 220-255. \\

Dawid, H., Gemkow, S., Harting, P., van der Hoog, S. and  Neugart, M. (2018). Agent-based macroeconomic modeling and policy analysis: the \textsc{eurace}@Unibi model. In: Chen, S.-H., Kaboudan, M., Du, Y.-R. (Eds.). \textit{The Oxford Handbook on Computational Economics and Finance}, Oxford University Press.\\

de Carvalho, F.J.C. (2015). Keynes on expectations, uncertainty and defensive behavior. \textit{Brazilian Keynesian Review}, vol.1(1), pp.44--54. \\

Delli Gatti, D., Gaffeo, E., Gallegati, M., Giulioni, G. and Palestrini, A. (2008). \textit{Emergent macroeconomics: an agent-based approach to business fluctuations}, Springer Science \& Business Media, Berlin. \\

Delli Gatti, D., Gallegati, M., Cirillo, P., Desiderio, S. and Gaffeo, E. (2011). \textit{Macroeconomics from the Bottom-Up}, Springer-Verlag, Berlin.\\

Delli Gatti, D., Di Guilmi, C., Gaffeo, E.,  Giulioni, G.,  Gallegati, M. and Palestrini, A. (2005). A new approach to business fluctuations: heterogeneous interacting agents, scaling laws and financial fragility. \textit{Journal of Economic behavior \& organization}, vol.56(4), pp.489--512. \\

Dillard, D. (1948). The Keynesian revolution and economic development. \textit{The Journal of Economic History}, vol.8(2), pp.171--177. \\

Dillard, D. (1980). A monetary theory of production: Keynes and the institutionalists. \textit{Journal of Economic Issues}, vol.14(2), pp.255--273. \\

Dosi, G., Fagiolo, G. and Roventini, A. (2006). An evolutionary model of endogenous business cycles. \textit{Computational Economics}, vol.27, pp.3--34. \\

Dosi, G., Faillo, M. and Marengo, L. (2008). Organizational capabilities, patterns of knowledge accumulation and governance structures in business firms: an introduction. \textit{Organization Studies}, vol.29(8-9), pp.1165--1185. \\

Dosi, G., Fagiolo, G. and Roventini, A. (2010). Schumpeter meeting Keynes: a policy-friendly model of endogenous growth and business cycles. \textit{Journal of Economic Dynamics and Control}, vol.34,  pp.1748–1767.\\

Dostaler, G. (2010). \textit{Keynes et ses combats}, Albin Michel, Paris. \\

Dow, S. and Hillard, J. (1995). \textit{Keynes, knowledge and uncertainty}, Edward Elgar Publishing, Cheltenham. \\

Epstein, J.M. (1999). Agent-based computational models and generative social science. \textit{Complexity}, vol.4(5), pp.41–60.\\

Gaffeo, E., Delli Gatti, D., Desiderio, S. and Gallegati M. (2008). Adaptive Microfoundations for Emergent Macroeconomics. \textit{Eastern Economic Journal}, vol.34(4), pp.441–463.  doi:10.1057/eej.2008.27 \\ 

Gallegati, M. and Richiardi, M.G. (2009). \textit{Agent Based Models in Economics and Complexity}. In Meyers, R. (eds), \textit{Encyclopedia of complexity and system science}, Springer, New York, pp.200--224. 	\\

Guerini M., Napoletano, M., Roventini, A. (2018). No man is an Island: The impact of heterogeneity and local interactions on macroeconomic dynamics, \textit{Economic Modelling}, 68, pp.82--95.\\

Harrod, R.F. (1937). Mr. Keynes and traditional theory. \textit{Econometrica}, vol.5(1), pp.74--86. \\

Harcourt, G. \textit{Some Cambridge Controversies in the Theory of Capital}. Cambridge: Cambridge University Press, 1972.\\

Hartley, R. I. (1997). In defense of the eight-point algorithm. \textit{IEEE Transactions on pattern analysis and machine intelligence}, vol.19(6), pp.580--593. \\

Hicks, J.R. (1937).  Mr. Keynes and the "Classics''. A suggested interpretation. \textit{Econometrica}, vol.5(2), pp.147-159.\\

Hicks, J.R. (1979). \textit{Causality in Economics}, Basic Books, New York.\\

Kahn, R. F. (1984). \textit{The making of Keynes' general theory}, Cambridge University Press, Cambridge.\\

Keynes, J.M. (1936). \textit{The General Theory of Unemployment, Interest, and Money}, MacMillan, London.\\

Kirman, A. (1992). Whom or What Does the Representative Individual Represent? \textit{Journal of Economic Perspectives}, vol.6(2), pp.117-136.\\

Kirman, A. (2011). \textit{Complex economics: individual and collective rationality}, Routledge, London. \\

Knight, F. H. (1921). Risk, uncertainty and profit (Vol. 31). Houghton Mifflin.\\

Kregel, J. A. (1976). Economic methodology in the face of uncertainty: the modelling methods of Keynes and the post-Keynesians. \textit{ The Economic Journal}, vol.86(342), pp.209--225. \\

Krugman, P. (2011). Mr. Keynes and the moderns. In \textit{Cambridge UK conference commemorating the 75th anniversary of the publication of The General theory of Employment, Interest, and Money}. \\

Ladyman, J.,  Lambert, J. and Wiesner, K. (2013). What is a complex system? \textit{European Journal for Philosophy of Science}, vol.3, pp.33--67. \\

Lawson, C.L. and Lawson, L.L. (1990). Financial system restructuring: lessons from Veblen, Keynes, and Kalecki. \textit{Journal of Economic Issues}, vol.24(1), pp.115--131. \\

Lawson, T. and Pesaran, H. (Eds.) (1985). \textit{Keynes' Economics: Methodological Issues}, Croom Helm, London. \\

Leijonhufvud, A. (1967).  Keynes and the Keynesians: A suggested interpretation. \textit{The American Economic Review}, vol.57(2), pp.401--410. \\

Leijonhufvud, A. (1968). Keynes and the effectiveness of monetary policy. \textit{Economic Inquiry}, vol.6(2), p.97.\\

Leijonhufvud, A. (2006). Agent-based Macro, in Tesfatsion L. and Judd K. (eds.). \textit{Handbook of Computational Economics}, vol.2, pp.1625-1637, Elsevier.\\

Leith, J. C. and Patinkin, D. (1977). \textit{Keynes, Cambridge and the General Theory}, Springer, New York. \\

Leontief, W. (1936). The fundamental assumption of Mr. Keynes' monetary theory of unemployment. \textit{The quarterly journal of economics}, vol.51(1), pp.192--197. \\

Majorana, E. (1942).  Il valore delle leggi statistiche nella fisica e nelle scienze sociali. \textit{Scientia}, vol.36, pp.58--66. \\

Massaro, D. (2013). Heterogeneous expectations in monetary DSGE models. \textit{Journal of Economic Dynamics \& Control} 37 680--692.\\

Meade, J. E. (1937). A simplified model of Mr. Keynes' system. \textit{The review of economic studies}, vol.4(2), pp.98--107. \\

Minsky, H.P. (1975). \textit{John Maynard Keynes}, Columbia University Press, New York.\\

Minsky, H. P. (1986). Money and crisis in Schumpeter and Keynes. \textit{The economic law of motion of modern society: A Marx-Keynes-Schumpeter centennial}, 112-122.

Moggridge, D. E. and Howson, S. (1974). Keynes on monetary policy, 1910--1946. \textit{Oxford Economic Papers}, vol.26(2), pp.226--247. \\

Moggridge, D. E. (1976). \textit{Keynes}, Springer, New York. \\

Nell, E. J.  (1998). \textit{The general theory of transformational growth}, Cambridge University Press, Cambridge. \\

Pasinetti, L. L. (1997). The Marginal Efficiency of Investment. In Harcourt, G., Riach, P. (ed.) \textit{A “Second Edition” of the General Theory}, Routledge, London, pp.198-218.  [http://hdl.handle.net/10807/80180]\\

Patinkin, D. (1990). On different interpretations of the general theory. \textit{Journal of Monetary Economics}, vol.26(2),  pp.205--243. \\

Pigou, A. C. (1936). Mr. J.M. Keynes' general theory of employment, interest and money. \textit{Economica}, vol.3(10), pp.115--132. \\

Raberto, M.,  Teglio, A. and Cincotti, S. (2014). Fiscal consolidation and sovereign debt risk in balance-sheet recessions. \textit{Economic Policy and the Financial Crisis}, pp.163, Routledge, London.\\

Robertson, D. H. (1938).  Mr. Keynes and “finance”. \textit {The Economic Journal}, vol.4(190), pp.314--318. \\

Robinson, J. The Production Function and the Theory oí Capital. \textit{Review of Economic studies}, Winter 1953-54, pp.81--106.\\

Robinson, E. (1964a). Could there have been a “general theory” without Keynes? \textit{Keynes’ General Theory: Reports of Three Decades}, pp.87--95. \\

Robinson, E. (1964b). John Maynard Keynes 1883--1946. \textit{Keynes’ General Theory: Reports of Three Decades}, pp.13--86. \\

Robinson, J. (1978).  Keynes and Ricardo. \textit{Journal of Post Keynesian Economics}, vol.1(1), pp.12--18. \\

Salle, I., Y{\i}ld{\i}zo{\u{g}}lu, M.  and  S{\'e}n{\'e}gas, M.A. (2013). Inflation targeting in a learning economy: An \textsc{abm} perspective. \textit{Economic Modelling}, vol.34, pp.114--128. \\

Salle, I. (2015).  Modeling expectations in agent-based models—an application to central bank's communication and monetary policy. \textit{Economic Modelling}, vol.46, pp.130--141.\\

Samuelson, P. A. (1946).  Lord Keynes and the general theory. \textit{Econometrica},  vol.14(3), pp.187--200. \\

Schumpeter, J. A. (1946). John Maynard Keynes 1883-1946. \textit{The American Economic Review}, vol.36(4), pp.495--518. \\

Seppecher, P. and Salle, I. (2015). Deleveraging crises and deep recessions: a behavioural approach. \textit{Applied Economics}, vol.47(34-35),  pp.3771--3790.  \\

Skidelsky, R. (2003). \textit{John Maynard Keynes: Economist, Philosopher, Statesman}. Macmillan.

Sornette, D. (2009). \textit{Why stock markets crash: critical events in complex financial systems}, Princeton University Press, Princeton. \\

Sraffa, P. (1960). \textit{Production of Commodities by Means of Commodities. Prelude to a critique of Economic Theory}, Cambridge University Press, Cambridge. \\

Tobin, J. (1977). How dead is Keynes? \textit{Economic Enquiry}, vol.15(4), pp.459--468. \\

Viner, J. (1936). Mr. Keynes on the causes of unemployment. \textit{The Quarterly Journal of Economics}, vol.51(1), pp.147--167. \\

Wolf, S., Bouchaud, J.P., Cecconi, F., Cincotti, S.,  Dawid, H., Gintis, H., van der Hoog, S., Jaeger, C.C., Kovalevsky, D.V. and  Mandel, A. (2013). Describing economic agent-based models--Dahlem \textsc{abm} documentation guidelines.
\textit{Complexity Economics}, vol.2(1), pp.63--74. \\

Wood, J.C. (1994) \textit{John Maynard Keynes, Critical Assessment: Second Series}, Psychology Press, vol.5. \\

Worswick, D., Worswick, G.D.N. and Trevithick, J. (1983). \textit{Keynes and the modern world}, CUP Archive.\\

\newpage
\section*{Appendix: simulation tables and validation}\label{app:simulation-results}

This appendix reports the tables underlying the figures in Section~\ref{sec:results}. Monetary magnitudes are expressed in wage units; when indicated, $Y^w$ is reported in millions of wage units.

\subsection*{A1. Accounting identities and multipliers}

\begin{table}[H]\scriptsize
\centering
\begin{tabular}{llrrrrrr}
\toprule
Regime & $h$ & $Y^w$ & $C^w$ & $I^w$ & $S^w-I^w$ & $\kappa$ & $Y^{w,*}-Y^w$ \\
\midrule
Homogeneous & 0.1\% & 4.726 & 1.792 & 2.934 & -0.00667 & 1.611 & 0.0107 \\
Homogeneous & 1.0\% & 4.726 & 1.792 & 2.934 & -0.00667 & 1.611 & 0.0107 \\
Homogeneous & 2.0\% & 4.726 & 1.792 & 2.934 & -0.00667 & 1.611 & 0.0107 \\
Homogeneous & 3.0\% & 4.726 & 1.792 & 2.934 & -0.00667 & 1.611 & 0.0107 \\
Homogeneous & 3.8\% & 0.293 & 0.000 & 0.293 & 0 & 1.000 & 0 \\
Homogeneous & 4.0\% & 0.293 & 0.000 & 0.293 & 0 & 1.000 & 0 \\
Homogeneous & 5.0\% & 0.293 & 0.000 & 0.293 & 0 & 1.000 & 0 \\
\addlinespace[0.25em]
Het. $AnSp$ & 0.1\% & 4.721 & 1.786 & 2.936 & -0.000667 & 1.608 & 0.00107 \\
Het. $AnSp$ & 1.0\% & 4.722 & 1.788 & 2.934 & -6.84e-05 & 1.609 & 0.00011 \\
Het. $AnSp$ & 2.0\% & 4.722 & 1.788 & 2.934 & -5.91e-05 & 1.609 & 9.51e-05 \\
Het. $AnSp$ & 3.0\% & 4.722 & 1.788 & 2.934 & 3.92e-05 & 1.609 & -6.31e-05 \\
Het. $AnSp$ & 3.8\% & 0.515 & 0.029 & 0.486 & 2.91e-10 & 1.059 & 0 \\
Het. $AnSp$ & 4.0\% & 0.292 & 0.000 & 0.292 & 0 & 1.000 & 0 \\
Het. $AnSp$ & 5.0\% & 0.293 & 0.000 & 0.293 & 0 & 1.000 & 0 \\
\addlinespace[0.25em]
Het. $L_2$ & 0.1\% & 4.726 & 1.792 & 2.934 & -0.00667 & 1.611 & 0.0107 \\
Het. $L_2$ & 1.0\% & 4.726 & 1.792 & 2.934 & -0.00667 & 1.611 & 0.0107 \\
Het. $L_2$ & 2.0\% & 4.726 & 1.792 & 2.934 & -0.00667 & 1.611 & 0.0107 \\
Het. $L_2$ & 3.0\% & 4.632 & 1.754 & 2.878 & -0.00503 & 1.609 & 0.0081 \\
Het. $L_2$ & 3.8\% & 1.326 & 0.335 & 0.991 & 1.48e-08 & 1.338 & -1.33e-08 \\
Het. $L_2$ & 4.0\% & 0.972 & 0.171 & 0.802 & 1.37e-08 & 1.213 & -4.19e-09 \\
Het. $L_2$ & 5.0\% & 0.275 & 0.001 & 0.274 & 2.91e-10 & 1.005 & 0 \\
\addlinespace[0.25em]
Het. $I^c$ & 0.1\% & 4.759 & 1.792 & 2.968 & -0.00667 & 1.604 & 0.0107 \\
Het. $I^c$ & 1.0\% & 3.934 & 1.500 & 2.434 & -0.000239 & 1.616 & 0.000387 \\
Het. $I^c$ & 2.0\% & 3.380 & 1.273 & 2.107 & -1.69e-05 & 1.604 & 2.71e-05 \\
Het. $I^c$ & 3.0\% & 2.892 & 0.988 & 1.904 & 1.72e-08 & 1.519 & -3.59e-08 \\
Het. $I^c$ & 3.8\% & 2.408 & 0.747 & 1.661 & 2.31e-08 & 1.450 & -7.92e-09 \\
Het. $I^c$ & 4.0\% & 2.229 & 0.646 & 1.583 & 3e-08 & 1.408 & -8.38e-09 \\
Het. $I^c$ & 5.0\% & 1.746 & 0.363 & 1.383 & 2.56e-08 & 1.262 & -7.45e-09 \\
\bottomrule
\end{tabular}
\caption{Accounting identity and multiplier verification by financial threshold.}
\label{tab:appendix-accounting-multiplier-threshold}
\end{table}

\begin{table}[H]\scriptsize
\centering
\begin{tabular}{lrrrrr}
\toprule
Case & $AnSp$ & $N_k$ & $N_c$ & $N_k+N_c$ & $\kappa_N^{\Delta}$ \\
\midrule
Homogeneous & 1 & 110.0 & 810.0 & 920.0 &  \\
Homogeneous & 2 & 230.0 & 1650.0 & 1880.0 & 8.00 \\
Homogeneous & 3 & 360.0 & 2520.0 & 2880.0 & 7.84 \\
Homogeneous & 4 & 495.0 & 3420.0 & 3915.0 & 7.78 \\
Homogeneous & 5 & 635.0 & 4345.0 & 4980.0 & 7.73 \\
\addlinespace[0.25em]
Monte Carlo & 1 & 110.6 & 824.0 & 934.6 &  \\
Monte Carlo & 2 & 231.1 & 1647.3 & 1878.4 & 7.83 \\
Monte Carlo & 3 & 360.8 & 2523.0 & 2883.8 & 7.79 \\
Monte Carlo & 4 & 495.9 & 3418.7 & 3914.5 & 7.73 \\
Monte Carlo & 5 & 633.5 & 4334.2 & 4967.8 & 7.71 \\
\bottomrule
\end{tabular}
\caption{Employment multiplier along the diagonal animal-spirits path. The multiplier is computed relative to $AnSp=1$.}
\label{tab:appendix-kahn-diagonal}
\end{table}

\subsection*{A2. Calibration of technology prices}

\begin{table}[H]\scriptsize
\centering
\begin{tabular}{rrrrrrrr}
\toprule
$I^c_c$ & $\overline{\mathrm{MEK}}_c$ & Interior & Floor & Ceiling & $C^+(1\%)$ & $C^+(3\%)$ & $C^+(5\%)$ \\
\midrule
$450,000$ & $5.47\%$ & $0.37$ & $0.00$ & $14.63$ & $14.96$ & $14.96$ & $14.89$ \\
$500,000$ & $5.15\%$ & $3.31$ & $0.04$ & $11.65$ & $14.89$ & $14.37$ & $12.11$ \\
$540,000$ & $4.35\%$ & $7.10$ & $0.50$ & $7.39$ & $14.30$ & $11.59$ & $8.15$ \\
$570,000$ & $3.49\%$ & $8.66$ & $1.82$ & $4.52$ & $12.44$ & $8.89$ & $4.63$ \\
$600,000$ & $2.54\%$ & $8.46$ & $4.22$ & $2.32$ & $10.00$ & $5.81$ & $2.59$ \\
$630,000$ & $1.81\%$ & $7.01$ & $6.91$ & $1.08$ & $7.26$ & $3.26$ & $1.04$ \\
$660,000$ & $1.26\%$ & $4.86$ & $9.57$ & $0.57$ & $4.41$ & $2.15$ & $0.67$ \\
$700,000$ & $0.85\%$ & $2.53$ & $12.26$ & $0.21$ & $2.07$ & $0.93$ & $0.26$ \\
$750,000$ & $0.60\%$ & $0.91$ & $14.05$ & $0.04$ & $0.63$ & $0.19$ & $0.04$ \\
\bottomrule
\end{tabular}
\caption{Calibration of $I^c_c$. $C^+(x)$ is the mean number of c-firms with $\mathrm{MEK}\ge x$.}
\label{tab:appendix-icc-calibration}
\end{table}

\begin{table}[H]\scriptsize
\centering
\begin{tabular}{rrrrrrr}
\toprule
$h$ & $C^+$ & Active C & $u$ & $Y^w$ & Mixed & Strong \\
\midrule
$0.05\%$ & $15.00$ & $15.00$ & $41.49\%$ & $4.203$ & $0.0\%$ & $0.0\%$ \\
$0.50\%$ & $15.00$ & $15.00$ & $41.62\%$ & $4.228$ & $0.0\%$ & $0.0\%$ \\
$1.00\%$ & $12.44$ & $12.44$ & $50.25\%$ & $3.469$ & $92.6\%$ & $92.6\%$ \\
$1.50\%$ & $12.07$ & $12.07$ & $52.03\%$ & $3.314$ & $100.0\%$ & $100.0\%$ \\
$2.00\%$ & $10.89$ & $10.89$ & $54.92\%$ & $3.032$ & $100.0\%$ & $100.0\%$ \\
$2.50\%$ & $10.30$ & $10.30$ & $57.68\%$ & $2.850$ & $96.3\%$ & $92.6\%$ \\
$3.00\%$ & $8.89$ & $8.89$ & $61.82\%$ & $2.481$ & $100.0\%$ & $96.3\%$ \\
$3.50\%$ & $8.59$ & $8.59$ & $63.28\%$ & $2.393$ & $100.0\%$ & $92.6\%$ \\
$4.00\%$ & $7.11$ & $7.11$ & $68.63\%$ & $1.994$ & $100.0\%$ & $96.3\%$ \\
$4.50\%$ & $5.96$ & $5.96$ & $71.26\%$ & $1.707$ & $100.0\%$ & $85.2\%$ \\
$5.00\%$ & $4.63$ & $4.63$ & $76.77\%$ & $1.363$ & $100.0\%$ & $77.8\%$ \\
\bottomrule
\end{tabular}
\caption{Selection curve for the central calibration $I^c_k=30{,}000$, $I^c_c=570{,}000$. $Y^w$ is reported in millions of wage units.}
\label{tab:appendix-selection-570000}
\end{table}

\subsection*{A3. Animal spirits and sectoral complementarity}

\begin{table}[H]\scriptsize
\centering
\begin{tabular}{rrrrrr}
\toprule
$AnSp_c\backslash AnSp_k$ & $1$ & $2$ & $3$ & $4$ & $5$ \\
\midrule
$5$ & $4.038$ & $4.260$ & $4.466$ & $4.655$ & $4.726$ \\
$4$ & $4.052$ & $4.215$ & $4.423$ & $4.463$ & $4.496$ \\
$3$ & $4.021$ & $4.185$ & $4.208$ & $4.240$ & $4.277$ \\
$2$ & $3.953$ & $3.962$ & $3.989$ & $4.024$ & $4.061$ \\
$1$ & $3.726$ & $3.750$ & $3.784$ & $3.819$ & $3.859$ \\
\bottomrule
\end{tabular}
\caption{Homogeneous animal-spirits matrix: income $Y^w$ in millions of wage units.}
\label{tab:appendix-ansp-homo-y}
\end{table}

\begin{table}[H]\scriptsize
\centering
\begin{tabular}{rrrrrr}
\toprule
$AnSp_c\backslash AnSp_k$ & $1$ & $2$ & $3$ & $4$ & $5$ \\
\midrule
$5$ & $55.42$ & $37.33$ & $20.64$ & $5.36$ & $0.00$ \\
$4$ & $54.28$ & $41.12$ & $24.20$ & $21.39$ & $19.18$ \\
$3$ & $56.85$ & $43.57$ & $42.17$ & $40.06$ & $37.55$ \\
$2$ & $62.55$ & $62.25$ & $60.54$ & $58.13$ & $55.62$ \\
$1$ & $81.53$ & $80.02$ & $77.71$ & $75.30$ & $72.49$ \\
\bottomrule
\end{tabular}
\caption{Homogeneous animal-spirits matrix: unemployment rate.}
\label{tab:appendix-ansp-homo-u}
\end{table}

\begin{table}[H]\scriptsize
\centering
\begin{tabular}{rrrrrr}
\toprule
$AnSp_c\backslash AnSp_k$ & $1$ & $2$ & $3$ & $4$ & $5$ \\
\midrule
$5$ & $3.989$ & $4.251$ & $4.427$ & $4.632$ & $4.723$ \\
$4$ & $4.049$ & $4.169$ & $4.401$ & $4.463$ & $4.497$ \\
$3$ & $4.018$ & $4.181$ & $4.209$ & $4.240$ & $4.275$ \\
$2$ & $3.952$ & $3.962$ & $3.990$ & $4.023$ & $4.062$ \\
$1$ & $3.729$ & $3.752$ & $3.786$ & $3.822$ & $3.861$ \\
\bottomrule
\end{tabular}
\caption{Monte Carlo animal-spirits matrix: mean income $Y^w$ in millions of wage units.}
\label{tab:appendix-ansp-mc-y}
\end{table}

\begin{table}[H]\scriptsize
\centering
\begin{tabular}{rrrrrr}
\toprule
$AnSp_c\backslash AnSp_k$ & $1$ & $2$ & $3$ & $4$ & $5$ \\
\midrule
$5$ & $56.09$ & $38.06$ & $23.90$ & $7.27$ & $0.24$ \\
$4$ & $54.51$ & $41.93$ & $26.06$ & $21.40$ & $19.10$ \\
$3$ & $57.07$ & $43.95$ & $42.09$ & $40.08$ & $37.68$ \\
$2$ & $62.63$ & $62.28$ & $60.47$ & $58.21$ & $55.54$ \\
$1$ & $81.23$ & $79.79$ & $77.50$ & $75.02$ & $72.37$ \\
\bottomrule
\end{tabular}
\caption{Monte Carlo animal-spirits matrix: mean unemployment rate.}
\label{tab:appendix-ansp-mc-u}
\end{table}

\begin{table}[H]\scriptsize
\centering
\begin{tabular}{rrrrrr}
\toprule
$AnSp_c\backslash AnSp_k$ & $1$ & $2$ & $3$ & $4$ & $5$ \\
\midrule
$5$ & $0.425$ & $0.621$ & $0.770$ & $0.940$ & $0.997$ \\
$4$ & $0.564$ & $0.748$ & $0.960$ & $1.001$ & $1.001$ \\
$3$ & $0.735$ & $0.994$ & $1.001$ & $1.001$ & $1.001$ \\
$2$ & $1.002$ & $1.001$ & $1.001$ & $1.001$ & $1.001$ \\
$1$ & $1.004$ & $1.003$ & $1.002$ & $1.002$ & $1.002$ \\
\bottomrule
\end{tabular}
\caption{Monte Carlo animal-spirits matrix: realized c-output relative to autonomous c-expectations.}
\label{tab:appendix-ansp-c-realization}
\end{table}

\begin{table}[H]\scriptsize
\centering
\begin{tabular}{rrrrrr}
\toprule
$AnSp_c\backslash AnSp_k$ & $1$ & $2$ & $3$ & $4$ & $5$ \\
\midrule
$5$ & $0.00$ & $0.00$ & $0.00$ & $0.00$ & $0.96$ \\
$4$ & $0.00$ & $0.00$ & $0.00$ & $0.00$ & $0.00$ \\
$3$ & $0.00$ & $0.00$ & $0.00$ & $0.00$ & $0.00$ \\
$2$ & $0.00$ & $0.00$ & $0.00$ & $0.00$ & $0.00$ \\
$1$ & $0.00$ & $0.00$ & $0.00$ & $0.00$ & $0.00$ \\
\bottomrule
\end{tabular}
\caption{Monte Carlo animal-spirits matrix: probability of unemployment below or equal to one percent.}
\label{tab:appendix-ansp-full-employment}
\end{table}

\subsection*{A4. Financial threshold and MEK selection}

\begin{table}[H]\scriptsize
\centering
\begin{tabular}{lrrrr}
\toprule
Regime & $h$ & Active C & $u$ & $Y^w$ \\
\midrule
homogeneous & $0.05\%$ & $15.00$ & $0.00\%$ & $4.726$ \\
homogeneous & $0.50\%$ & $15.00$ & $0.00\%$ & $4.726$ \\
homogeneous & $1.00\%$ & $15.00$ & $0.00\%$ & $4.726$ \\
homogeneous & $2.00\%$ & $15.00$ & $0.00\%$ & $4.726$ \\
homogeneous & $3.00\%$ & $15.00$ & $0.00\%$ & $4.726$ \\
homogeneous & $3.80\%$ & $0.00$ & $87.25\%$ & $0.293$ \\
homogeneous & $4.00\%$ & $0.00$ & $87.25\%$ & $0.293$ \\
homogeneous & $5.00\%$ & $0.00$ & $87.25\%$ & $0.293$ \\
het. AnSp & $0.05\%$ & $15.00$ & $0.35\%$ & $4.721$ \\
het. AnSp & $0.50\%$ & $15.00$ & $0.26\%$ & $4.722$ \\
het. AnSp & $1.00\%$ & $15.00$ & $0.29\%$ & $4.722$ \\
het. AnSp & $2.00\%$ & $15.00$ & $0.26\%$ & $4.722$ \\
het. AnSp & $3.00\%$ & $15.00$ & $0.27\%$ & $4.722$ \\
het. AnSp & $3.80\%$ & $0.75$ & $82.88\%$ & $0.515$ \\
het. AnSp & $4.00\%$ & $0.00$ & $87.27\%$ & $0.292$ \\
het. AnSp & $5.00\%$ & $0.00$ & $87.25\%$ & $0.293$ \\
het. $L_2$ & $0.05\%$ & $15.00$ & $0.00\%$ & $4.726$ \\
het. $L_2$ & $0.50\%$ & $15.00$ & $0.00\%$ & $4.726$ \\
het. $L_2$ & $1.00\%$ & $15.00$ & $0.00\%$ & $4.726$ \\
het. $L_2$ & $2.00\%$ & $15.00$ & $0.00\%$ & $4.726$ \\
het. $L_2$ & $3.00\%$ & $14.68$ & $1.82\%$ & $4.632$ \\
het. $L_2$ & $3.80\%$ & $3.49$ & $66.90\%$ & $1.326$ \\
het. $L_2$ & $4.00\%$ & $2.30$ & $73.87\%$ & $0.972$ \\
het. $L_2$ & $5.00\%$ & $0.06$ & $88.41\%$ & $0.275$ \\
het. $I^c$ & $0.05\%$ & $15.00$ & $0.00\%$ & $4.759$ \\
het. $I^c$ & $0.50\%$ & $14.82$ & $1.04\%$ & $4.667$ \\
het. $I^c$ & $1.00\%$ & $12.55$ & $14.18\%$ & $3.934$ \\
het. $I^c$ & $2.00\%$ & $10.76$ & $24.59\%$ & $3.380$ \\
het. $I^c$ & $3.00\%$ & $9.16$ & $33.89\%$ & $2.892$ \\
het. $I^c$ & $3.80\%$ & $7.54$ & $43.34\%$ & $2.408$ \\
het. $I^c$ & $4.00\%$ & $6.93$ & $46.92\%$ & $2.229$ \\
het. $I^c$ & $5.00\%$ & $5.28$ & $56.52\%$ & $1.746$ \\
\bottomrule
\end{tabular}
\caption{Financial threshold summary by regime. $Y^w$ is reported in millions of wage units.}
\label{tab:appendix-threshold-summary}
\end{table}

\subsection*{A5. Sections of the $r\times L_2$ map}

\begin{table}[H]\scriptsize
\centering
\begin{tabular}{rrrrrrr}
\toprule
$r$ & $L_2$ & $h$ & Active C & $u$ & $Y^w$ & $\lambda$ \\
\midrule
$1.00\%$ & $0.05\%$ & $1.00\%$ & $10.80$ & $24.36\%$ & $3.421$ & $0.000$ \\
$1.00\%$ & $0.50\%$ & $1.00\%$ & $11.80$ & $18.53\%$ & $3.723$ & $0.000$ \\
$1.00\%$ & $1.00\%$ & $1.00\%$ & $12.40$ & $15.04\%$ & $3.903$ & $0.500$ \\
$1.00\%$ & $2.00\%$ & $2.00\%$ & $11.80$ & $18.53\%$ & $3.691$ & $1.000$ \\
$1.00\%$ & $3.00\%$ & $3.00\%$ & $8.20$ & $39.50\%$ & $2.618$ & $1.000$ \\
$1.00\%$ & $4.00\%$ & $4.00\%$ & $7.20$ & $45.32\%$ & $2.291$ & $1.000$ \\
$1.00\%$ & $5.00\%$ & $5.00\%$ & $6.00$ & $52.31\%$ & $1.971$ & $1.000$ \\
\addlinespace[0.2em]
$3.00\%$ & $0.05\%$ & $3.00\%$ & $7.60$ & $42.99\%$ & $2.443$ & $0.000$ \\
$3.00\%$ & $0.50\%$ & $3.00\%$ & $8.00$ & $40.66\%$ & $2.565$ & $0.000$ \\
$3.00\%$ & $1.00\%$ & $3.00\%$ & $9.80$ & $30.18\%$ & $3.077$ & $0.000$ \\
$3.00\%$ & $2.00\%$ & $3.00\%$ & $10.40$ & $26.69\%$ & $3.243$ & $0.000$ \\
$3.00\%$ & $3.00\%$ & $3.00\%$ & $8.40$ & $38.33\%$ & $2.685$ & $0.500$ \\
$3.00\%$ & $4.00\%$ & $4.00\%$ & $6.60$ & $48.82\%$ & $2.134$ & $1.000$ \\
$3.00\%$ & $5.00\%$ & $5.00\%$ & $4.00$ & $63.96\%$ & $1.396$ & $1.000$ \\
\addlinespace[0.2em]
$5.00\%$ & $0.05\%$ & $5.00\%$ & $5.00$ & $58.13\%$ & $1.676$ & $0.000$ \\
$5.00\%$ & $0.50\%$ & $5.00\%$ & $5.60$ & $54.64\%$ & $1.834$ & $0.000$ \\
$5.00\%$ & $1.00\%$ & $5.00\%$ & $4.80$ & $59.30\%$ & $1.613$ & $0.000$ \\
$5.00\%$ & $2.00\%$ & $5.00\%$ & $5.60$ & $54.64\%$ & $1.815$ & $0.000$ \\
$5.00\%$ & $3.00\%$ & $5.00\%$ & $3.80$ & $65.12\%$ & $1.336$ & $0.000$ \\
$5.00\%$ & $4.00\%$ & $5.00\%$ & $5.60$ & $54.64\%$ & $1.836$ & $0.000$ \\
$5.00\%$ & $5.00\%$ & $5.00\%$ & $6.20$ & $51.14\%$ & $1.983$ & $0.500$ \\
\addlinespace[0.2em]
\bottomrule
\end{tabular}
\caption{$r$--$L_2$ sections for the Het. $I^c$ regime. $Y^w$ is reported in millions of wage units.}
\label{tab:appendix-r-lp-sections-r-lp-map-ic-mc}
\end{table}

\begin{table}[H]\scriptsize
\centering
\begin{tabular}{rrrrrrr}
\toprule
$r$ & $L_2$ & $h$ & Active C & $u$ & $Y^w$ & $\lambda$ \\
\midrule
$1.00\%$ & $0.05\%$ & $1.00\%$ & $15.00$ & $0.00\%$ & $4.726$ & $0.000$ \\
$1.00\%$ & $0.50\%$ & $1.00\%$ & $15.00$ & $0.00\%$ & $4.726$ & $0.000$ \\
$1.00\%$ & $1.00\%$ & $1.00\%$ & $15.00$ & $0.00\%$ & $4.726$ & $0.537$ \\
$1.00\%$ & $2.00\%$ & $2.00\%$ & $15.00$ & $0.00\%$ & $4.726$ & $1.000$ \\
$1.00\%$ & $3.00\%$ & $3.00\%$ & $13.60$ & $8.05\%$ & $4.313$ & $1.000$ \\
$1.00\%$ & $4.00\%$ & $4.00\%$ & $5.00$ & $58.13\%$ & $1.771$ & $1.000$ \\
$1.00\%$ & $5.00\%$ & $5.00\%$ & $0.40$ & $88.49\%$ & $0.329$ & $1.000$ \\
\addlinespace[0.2em]
$3.00\%$ & $0.05\%$ & $3.00\%$ & $15.00$ & $0.00\%$ & $4.726$ & $0.000$ \\
$3.00\%$ & $0.50\%$ & $3.00\%$ & $15.00$ & $0.00\%$ & $4.726$ & $0.000$ \\
$3.00\%$ & $1.00\%$ & $3.00\%$ & $15.00$ & $0.00\%$ & $4.726$ & $0.000$ \\
$3.00\%$ & $2.00\%$ & $3.00\%$ & $15.00$ & $0.00\%$ & $4.726$ & $0.000$ \\
$3.00\%$ & $3.00\%$ & $3.00\%$ & $14.20$ & $4.60\%$ & $4.490$ & $0.490$ \\
$3.00\%$ & $4.00\%$ & $4.00\%$ & $2.80$ & $70.94\%$ & $1.120$ & $0.959$ \\
$3.00\%$ & $5.00\%$ & $5.00\%$ & $0.40$ & $86.96\%$ & $0.364$ & $0.998$ \\
\addlinespace[0.2em]
$5.00\%$ & $0.05\%$ & $5.00\%$ & $0.00$ & $87.25\%$ & $0.293$ & $0.000$ \\
$5.00\%$ & $0.50\%$ & $5.00\%$ & $0.00$ & $87.25\%$ & $0.293$ & $0.000$ \\
$5.00\%$ & $1.00\%$ & $5.00\%$ & $0.00$ & $87.25\%$ & $0.293$ & $0.000$ \\
$5.00\%$ & $2.00\%$ & $5.00\%$ & $0.00$ & $87.25\%$ & $0.293$ & $0.000$ \\
$5.00\%$ & $3.00\%$ & $5.00\%$ & $0.00$ & $87.25\%$ & $0.293$ & $0.000$ \\
$5.00\%$ & $4.00\%$ & $5.00\%$ & $0.00$ & $87.25\%$ & $0.293$ & $0.050$ \\
$5.00\%$ & $5.00\%$ & $5.00\%$ & $0.00$ & $90.82\%$ & $0.211$ & $0.455$ \\
\addlinespace[0.2em]
\bottomrule
\end{tabular}
\caption{$r$--$L_2$ sections for the Het. $L_2$ regime. $Y^w$ is reported in millions of wage units.}
\label{tab:appendix-r-lp-sections-r-lp-map-lp-mc}
\end{table}

\subsubsection*{A51. Full \(r\times L_2\) unemployment maps}

\begin{figure}[H]
	\centering
	\includegraphics[width=.6\textwidth]{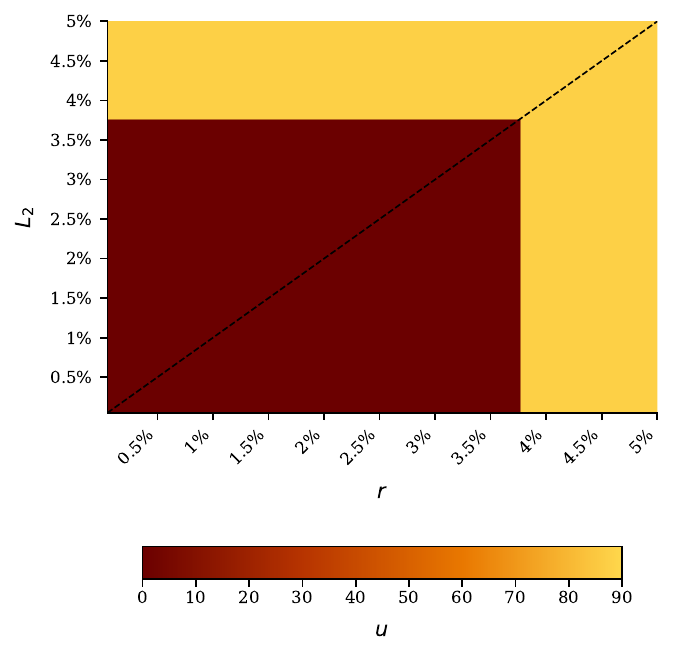}
	\caption{Unemployment in the \(r\times L_2\) plane: homogeneous benchmark.}
	\label{fig:app-homo-u-map}
\end{figure}

\begin{figure}[H]
	\centering
	\includegraphics[width=.6\textwidth]{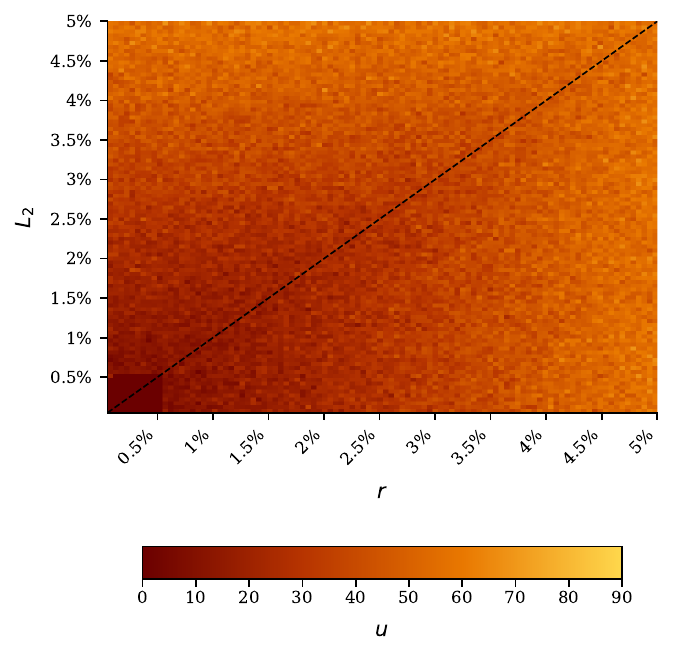}
	\caption{Unemployment in the \(r\times L_2\) plane: heterogeneous technology prices.}
	\label{fig:app-ic-u-map}
\end{figure}

\begin{figure}[H]
	\centering
	\includegraphics[width=.6\textwidth]{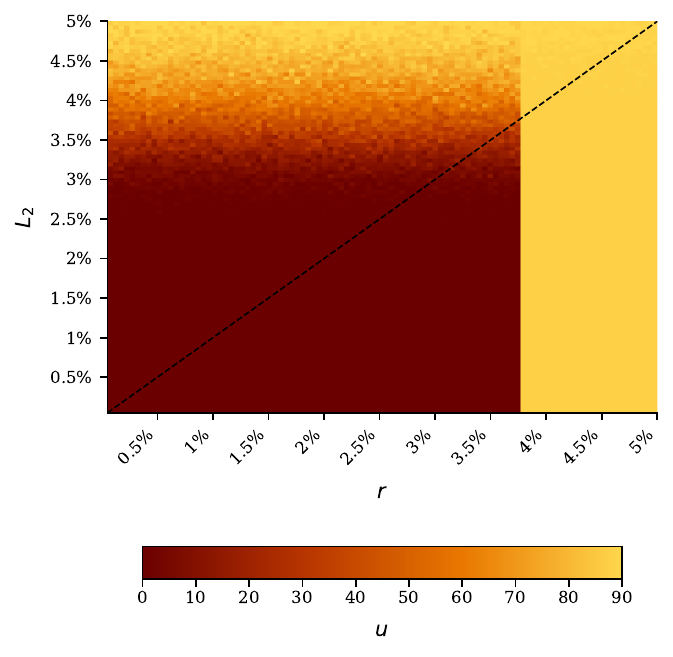}
	\caption{Unemployment in the \(r\times L_2\) plane: heterogeneous liquidity preferences.}
	\label{fig:app-lp-u-map}
\end{figure}

Figures~\ref{fig:app-homo-u-map}--\ref{fig:app-lp-u-map} report the full unemployment maps in the \(r\times L_2\) plane for the homogeneous benchmark, the heterogeneous technology-price case, and the heterogeneous liquidity-preference case.

\subsection*{A6. Validation}

\begin{table}[H]\scriptsize
\centering
\begin{tabular}{lllll}
\toprule
Experiment & Coverage & Errors & Accounting & Individual controls \\
\midrule
Phase 2 calibration & 891 grid points, 2,673 runs & 0 & 100\% & 100\% \\
$AnSp_k\times AnSp_c$ & 25 homogeneous cells and 1,250 MC runs & 0 & 100\% & 100\% \\
$r\times L_2$ maps & 40 run files, 160,000 run rows & 0 & 100\% & 100\% \\
\bottomrule
\end{tabular}
\caption{Validation of the simulation batches used in the final Results section.}
\label{tab:appendix-validation}
\end{table}

\end{document}